\documentclass[11pt]{article}

\usepackage{amssymb,amsmath,latexsym, hyperref, cite, natbib, graphicx, refstyle, pgfplotstable, booktabs, rotating, color, tabularx, authblk, array, url, soul, longtable,lscape,afterpage,cclicenses}

\usepackage{multirow}
\newcolumntype{C}[1]{>{\centering\arraybackslash}p{#1}}

\usepackage{geometry}
\usepackage{setspace}
\usepackage[multiple, hang,splitrule]{footmisc}


\usepackage[font=small, margin=0cm]{caption}

\begin{document}
\title{\textbf{A Triple Helix Model of Medical Innovation: \linebreak \textit{Supply},  \textit{Demand}, and  \textit{Technological Capabilities} in terms of Medical Subject Headings 
}}
\author[1]{\textbf{Alexander M. Petersen}\thanks{Corresponding author: petersen.xander@gmail.com}}
\author[2]{\textbf{Daniele Rotolo}\thanks{d.rotolo@sussex.ac.uk}}
\author[3]{\textbf{Loet Leydesdorff}\thanks{Corresponding author: loet@leydesdorff.net}}

\affil[1]{\footnotesize Laboratory for the Analysis of Complex Economic Systems, IMT Institute for Advanced Studies, Lucca, Italy}
\affil[2]{\footnotesize SPRU --- Science Policy Research Unit, University of Sussex, Brighton, United Kingdom}
\affil[3]{\footnotesize Amsterdam School of Communication Research (ASCoR), University of Amsterdam, Amsterdam, Netherlands}

\date{\small Version: \today \linebreak
Accepted for publication in \textit{\textbf{Research Policy}}\thanks{\href{http://dx.doi.org/10.1016/j.respol.2015.12.004}{{\color{blue}DOI: 10.1016/j.respol.2015.12.004}}. \copyright 2015 Petersen, Rotolo, Leydesdorff. Distributed under \href{http://creativecommons.org/licenses/by/4.0/}{{\color{blue}CC-BY-NC-ND}}.}}

\maketitle
\begin{abstract}
\singlespacing
\noindent \footnotesize We develop a model of innovation that enables us to trace the interplay among three key dimensions of the innovation process: (i) \textit{demand} of and (ii) \textit{supply} for innovation, and (iii) \textit{technological capabilities} available to generate innovation in the forms of products, processes, and services. Building on triple helix research, we use entropy statistics to elaborate an indicator of mutual information among these dimensions that can provide indication of reduction of uncertainty. To do so, we focus on the medical context, where uncertainty poses significant challenges to the governance of innovation. We use the Medical Subject Headings (MeSH) of MEDLINE/PubMed to identify publications classified within the categories ``Diseases'' ($C$), ``Drugs and Chemicals'' ($D$), ``Analytic, Diagnostic, and Therapeutic Techniques and Equipment'' ($E$) and use these as knowledge representations of \textit{demand}, \textit{supply}, and \textit{technological capabilities}, respectively. Three case-studies of medical research areas are used as representative 'entry perspectives' of the medical innovation process. These are: (i) human papilloma virus, (ii) RNA interference, and (iii) magnetic resonance imaging. We find statistically significant periods of synergy among \textit{demand}, \textit{supply}, and \textit{technological capabilities} ($C-D-E$) that point to three-dimensional interactions as a fundamental perspective for the understanding and governance of the uncertainty associated with medical innovation. Among the pairwise configurations in these contexts, the \textit{demand-technological capabilities} ($C-E$) provided the strongest link, followed by the \textit{supply-demand} ($D-C$) and the \textit{supply-technological capabilities} ($D-E$) channels. 
  \par
\end{abstract}
\clearpage 

\section{Introduction}
The development of models of innovation capable of increasing our understanding of the innovation process and of tracing/predicting innovation dynamics have been a longstanding central topic in the science-policy and innovation-studies literature as well as of policy debates \citep{Martin2012a}. The complexity of the models of innovation proposed has increased over time: the ``chain-linked'' model of innovation, for example, advanced on linear models (\textit{technology-push} and \textit{demand-pull}) by introducing feedback and feed forward loops among the different stages of the innovation process \citep{Kline1986}. However, such interactive models are not sufficient to explain what drives innovation and technological development and why certain firms are more capable than others in pursuing innovation \citep{Marinova2003}. Evolutionary economists, building on nonlinear feedback analysis from evolutionary biology, have instead pointed to the role of \textit{routines} (i.e., standardized patterns of actions representing 'genes') that firms use to develop products and services (along technological trajectories), which, in turn, generate variation \citep{Nelson1977, Nelson1982a}. Products and services compete in market and non-market \textit{selection environments} \citep{Nelson1977} including technological \citep{Dosi1982} and technoeconomic paradigms \citep{Perez1983}.

In such a framework, one can expect more than a single selection mechanism to be relevant in the case of innovation. In his study of post-Schumpeterian contributions, \cite[][p. 195]{Andersen1994} noted that ``(E)volutionary economics cannot rely on a standard form of explanation to the same extent as evolutionary biology.'' Biological evolution theory, assumes \textit{variation} as a driver and \textit{selection} to be naturally given, while cultural evolution is driven by individuals and groups who make conscious decisions on the basis of potentially different criteria \citep{Newell1972}. As such, the evolving construct is not a given unit of analysis \citep[][p. 14]{Andersen1992}.  Boulding \citeyearpar[][p. 33]{Boulding1978} suggested that ``(W)hat evolves is something very much like knowledge.'' Yet, not only bodies of knowledge are evolving, but also markets. Henceforth arises the basic question, under which conditions can the different selection mechanisms be expected to co-evolve and lead to (options for) new innovations? 

When different selection mechanisms can operate upon one another, a complex systems dynamic is generated. From this perspective, the model of the National Innovation System \citep[e.g.][]{Freeman1987,Lundvall1988,Nelson1993a}, and its subsequent extensions to systems of regional \citep{Braczyk1998} or sectorial \citep{Malerba2002} innovation, can be considered as the specification of possible levels of integration \citep[cf.][]{Carlsson2006}; but both integration and differentiation among selection environments can be expected to operate continuously in complex systems of innovation. The interactions among selection mechanisms generate options for innovation by decoding and recoding the relevant criteria \citep{Cowan1997} or, in other words, puzzle-solving \citep{Arthur2009a,Bradshaw1991}. 

The literature on the co-evolution between two selection environments highlights processes of mutual shaping \citep{McLuhan1964}, niche formation \citep{Schot2007}, or lock-in \citep{Arthur1989}. While stable equilibria are often attractors along the evolutionary pathway, pathways along trajectories can, however, become meta-stable or selected for globalization at the regime level, when three selection/variation mechanisms operate upon one another \citep{Etzkowitz2000}. Co-evolutions between two sub-dynamics can be continuously upset by a third, leading to crises, hyper-stability, and other complex phenomena \citep{Leydesdorff1998a,Ulanowicz2009}. 

\begin{figure}
\includegraphics[width=10cm]{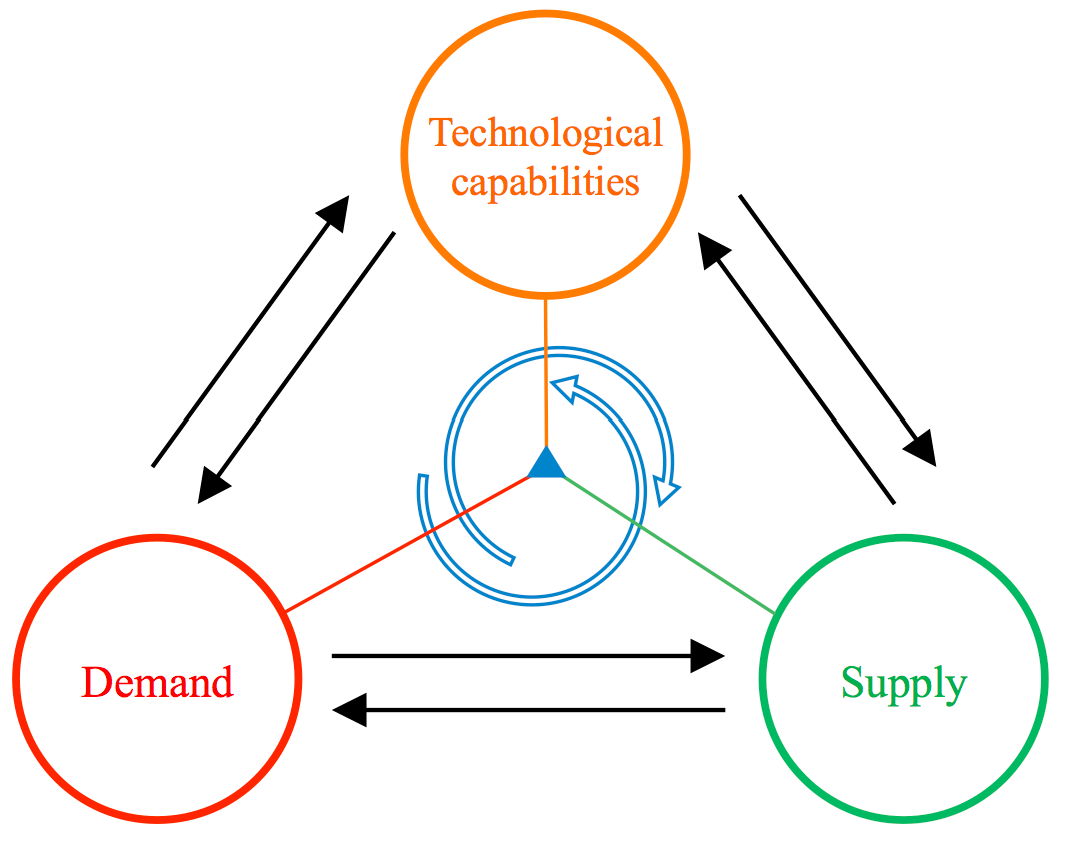}
\centering
\caption{Models of feedback loops based on interactions among: \textit{supply}, \textit{demand}, and \textit{technical capabilities}. The directionality of the arrows represents the possibility of differential strength in opposite directions. \newline\textit{Source: Authors' elaboration.}}
\label{fig:model}
\end{figure}

Here we consider a nonlinear three-dimensional model of innovation, with a specific focus on the medical context as discussed below. \Figref{model} depicts the interactions among three key dimensions in innovation studies: \textit{supply}-side factors, \textit{demand} articulation, and \textit{technological capabilities} (e.g.\ state-of-the-art instrumentations) to generate new products, processes and services. The triangle of arrows allows for --- potentially alternating --- clockwise and counter-clockwise rotations and even (next-order) loops. The relation between any two dynamics can be spuriously correlated upon by a third factor, which may enhance or dampen the relation between the other two. 

For example, the relation between \textit{demand} articulation and \textit{technological capabilities} may lead to new \textit{supply}-side offering of products, processes or services. In other words, the relations between each two dimensions can be auto-catalyzed by the third so that proliferations or extinctions become possible when the order of the arrows can circularly be closed into recursive loops \citep{Krippendorff2009a,Ulanowicz2009}. A self-organizing complex system thus can be expected to emerge from linear flows when feedback loops continue to exist \citep{Maturana2000}.

For the measurement of these complex dynamics, we turn to entropy statistics \citep{Shannon1948,Theil1972}. These measures have been used in triple helix research to build an indicator of mutual information (relational dependence) among three dimensions $x$, $y$, and $z$, namely $T_{xyz}$ \citep{McGill1954, Yeung2008} --- the mathematical formulation of this indicator will be provided and utilized as part of our analyses. Negative $T_{xyz}$ values have been associated with the reduction of the uncertainty that prevails at the system level because of synergetic integration, while positive values can be considered as indicating differentiation among the interactions \citep{Leydesdorff2014a}.\footnote{Unlike variance analysis, uncertainty analysis in terms of bits of information does not presume normality in the distributions \citep{Garner1956}.} \cite{Leydesdorff2014b} showed that negative information in a triple helix configuration finds its origin in redundancy that is generated when uncertainty is selected from different perspectives. New options are generated in the interactions among selection mechanisms. The total number of options --- the maximum entropy --- is thus increased. The increase in the redundancy may outweigh the increase of uncertainty generated in ongoing processes of variation.

The relevance of this indicator for innovation studies can be appreciated from the two perspectives of reducing uncertainty or increasing redundancy. First, one can expect a configuration with less uncertainty to be more rewarding with regards to risk-taking than periods with high uncertainty in the relevant (selection) environments. Reduction of the prevailing uncertainty provides innovators with dynamic opportunities comparable to local niches \citep[e.g.][]{Schot2007}. Note that reduction of uncertainty at the systems level provides an advantage for reflexive agency insofar as it is perceived. 

Second, the increase in redundancy itself is a structural effect at the systems level --- that is, a result of interacting selection mechanisms. The relative reduction of uncertainty in the configuration is caused by an increase of the redundancy in terms of the number of options available for innovation --- the two components (relative information and redundancy) are each other's complement, adding up to the maximum entropy of a system. Among the total number of options possible, the redundancy represents the configurations which have not (yet) been realized. An increase in this number does not necessarily affect the number of the realized options as long as the maximum number of options also increases \citep{Brooks1986,Khalil1996}. 

The number of options available to an innovation system for realization may be as decisive for its survival as the historically already-realized innovations. Although uncertainty features in all innovation processes \citep{Freeman1997b}, it poses crucial challenges to the governance of innovation especially in the medical context \citep{Consoli2016,Gelijns2001}, which is the focus of our analyses. Also, the importance of the interplay among \textit{supply}, \textit{demand}, and \textit{technological capabilities} in the medical innovation process is discussed in the framework on the progress of medical knowledge and practice proposed by \cite{Nelson2011} in terms of three enabling forces: advances of scientific understanding of diseases, learning in clinical practices, and advances in technological capabilities (which often originate outside of medicine) for the development of novel modalities of diagnosis and treatment. 

The Medical Subject Headings (MeSH) provided by the MEDLINE/PubMed publication data of the U.S. National Library of Medicine offer a valuable framework for operationalizing our research question about windows of opportunities in terms of numbers of options and possible reduction of the prevailing uncertainty in relation to the three sources identified by \cite{Nelson2011}.\footnote{The of US National Library of Medicine of the United States (NLM) have constantly received relatively large funding for maintaining and updating its biomedical and health information services --- for example, the 2015 budget for these services was of \$117 Million \citep{NLM2015}. This has enabled a relatively uniform application of the MeSH classification to publications by indexers over many years.} The MeSH classification provides a rich controlled vocabulary, composed of individual ``descriptors'' ($m_i$) that classify the topics/concepts of publications at different levels of specificity. The descriptors are organized in a tree-like network structure consisting of 16 branches (high-level topics) denoted by $\alpha=\{A, B, C, D, E, F, G, H, I, J, K, L, M, N, V, Z\}$. Within each branch, descriptors with shorter ``Tree Number'' identification codes are relatively general concepts that branch out into more specific concepts.\footnote{For example, the descriptor ``Nervous System'' [Tree Number: A08] includes several child descriptors of increasing specificity: ``Central Nervous System'' [Tree Number: A08.186], ``Brain'' [Tree Number: A08.186.211], ..., ``Pituitary Gland, Anterior'' [Tree Number: A08.186.211.464.497.352.435.500.500].}

\begin{figure}
\includegraphics[width=16cm]{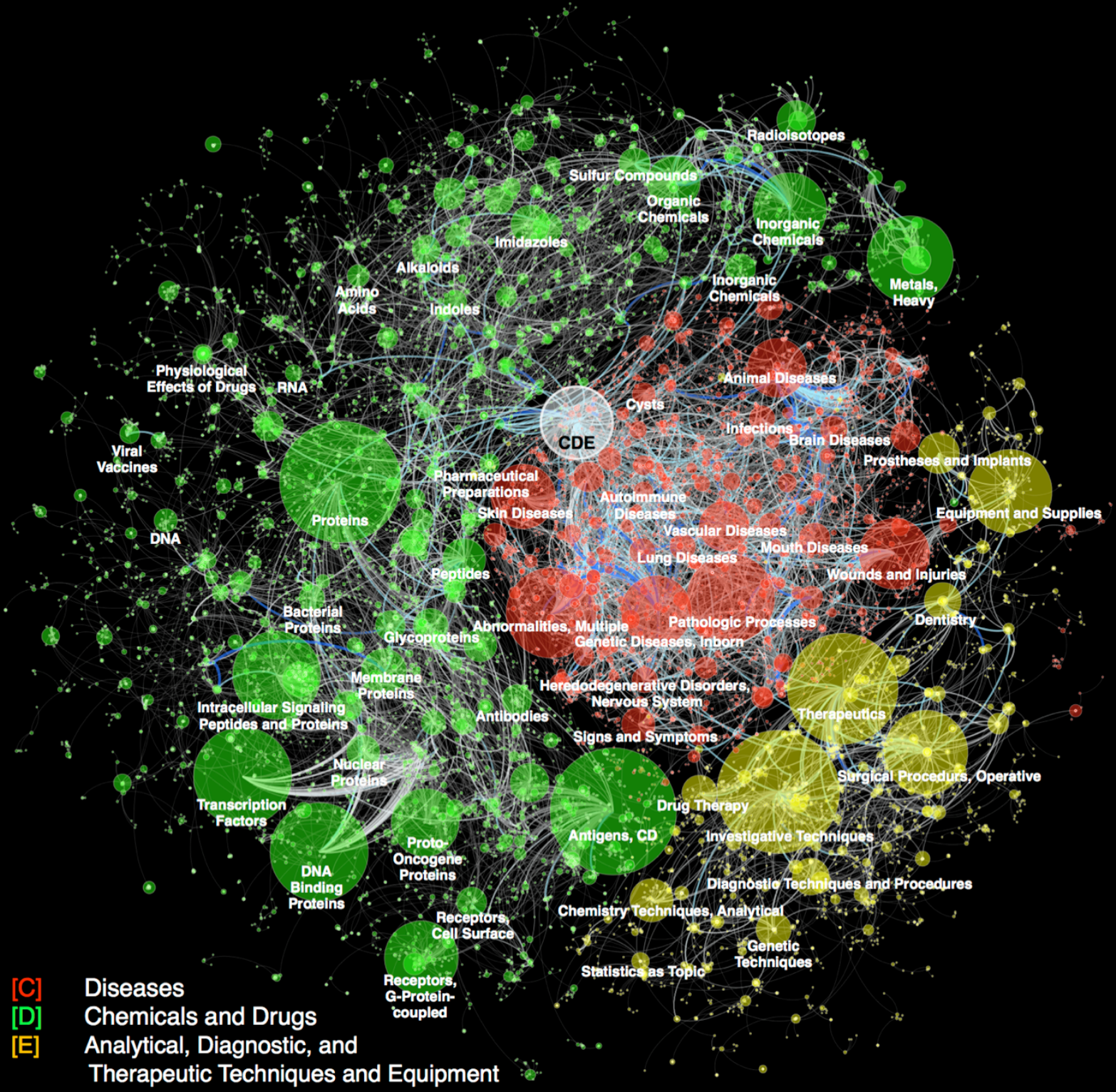}
\centering
\caption{Network visualization of the \textit{C}, \textit{D}, and \textit{E} branches ($\alpha$) of the 2014 MeSH classification. The network comprises 16,579 descriptors (nodes) and 23,475 links representing conceptual relations.. The 'CDE' node represents an artificial node that connects the first-level MeSH descriptors (e.g., C22, ``Animal Diseases''), used primarily for network layout purposes. The size of each node $i$ is nonlinearly proportional to its in-degree $k_i^{in}$ representing the number of associated descriptors that are immediately below $i$ in the classification tree; the node color corresponds to the main branch $\alpha$. The links are directed, representing the path from descriptor $j$ to $i$ in the direction of the main tree root 'CDE'. The thickness of each directed link is proportional to the total degree $k_i$, and the color is nonlinearly proportional to the in-degree divided by the (largest associated) branch level $L_j$ corresponding to the ratio $q_{ij} = k_i^{in}/L_j$. Links with exceptionally large $q_{ij}$ values, representing core MeSH associations, are colored more blue. \newline\textit{Source: Authors' elaboration on the basis of MEDLINE/PubMed data.}}
\label{fig:meshmap}
\end{figure}

On the basis of this classification and in the vein of previous studies \citep{Agarwal2008,Agarwal2009,Leydesdorff2012, Shi2015}, we use three MeSH branches as knowledge representations of \textit{supply}, \textit{demand}, and \textit{technological capabilities}. The ``Diseases'' branch ($\alpha = C$) is considered as a knowledge representation of \textit{demand} for innovations --- knowledge and understanding of diseases incentivizes both patients and doctors to articulate demand; the ``Drugs and Chemicals'' branch ($\alpha = D$) as a knowledge representation of the \textit{supply} side in terms of therapeutics diagnostics (e.g.\ biological markers); and the ``Analytic, Diagnostic, and Therapeutic Techniques and Equipment'' branch ($\alpha = E$) as a knowledge representation of the state of art of \textit{technological capabilities} (e.g.\ surgical procedures, investigative techniques) that can mediate the supply-demand nexus. We note that a single MeSH term $m_i$ is not intended to individually represent \textit{demand}, \textit{supply}, or \textit{technological capabilities}. Rather, we constructively use the co-occurrence statistics derived from the sets of $m_i$ occurring within each publication to project each research area onto the $C-D-E$ representation. \Figref{meshmap} shows the network structure of the MeSH system, indicating the breadth and depth of the knowledge domains represented by the three selected MeSH branches.

Our empirical strategy is structured as follows. We first explore the co-evolutionary dynamics within the $C$, $D$, and $E$ knowledge spaces and search for the presence of statistical regularities that enable us to characterize the MeSH classification system. To do so, we apply statistical formulations from quantitative linguistics and information theory. We then analyze the redundancy between the $C$, $D$, and $E$ as ``information channels'' first bi-laterally and then tri-laterally. The latter will provide indication on the emergence of synergies among \textit{supply}, \textit{demand}, and \textit{technological capabilities} over time arising from configurational feedback loops. We perform our analyses on the co-occurrence statistics of individual MeSH descriptors drawn from roughly 100,000 medical publications associated with three case-studies of medical areas characterized by breakthrough scientific discoveries and technological innovations, which are representative examples of \textit{supply}, \textit{demand}, and \textit{technological capabilities}. We use this framework to test the null hypothesis that the tri-lateral relations are fully specified by the superposition of purely bi-lateral interactions, which in terms of the schematic \Figref{model}, tests the presence of tri-lateral interactions represented by the circulating arrows.

\section{Case-studies and data}
To address our research question about innovation opportunities captured by the number of options and the possible reduction of the prevailing uncertainty among \textit{supply-demand-technological capabilities} in the medical context, we first identify case-studies of medical areas that (i) are representative of the three dimensions of our model as operationalized in terms of $C$, $D$, and $E$ branches, i.e. medical areas that provide three different 'entry perspectives' on the medical innovation process; (ii) can be characterised by prominent innovation dynamics in terms of breakthrough scientific discoveries and technological innovations; and (iii) for which qualitative insights that can support the meaningful interpretation of the results of our analyses are already available. 

Using these criteria, we selected the following case-studies: human papilloma virus (HPV), RNA interference (RNAi), and magnetic resonance imaging (MRI). These are considered as representative of $C$ (\textit{demand}), $D$ (\textit{supply}), and $E$ (\textit{technological capabilities}), respectively. Furthermore, the scientific discoveries and technological advancements within these medical areas were recognized by the award of Nobel Prizes in Physiology or Medicine. Recent research \citep{Leydesdorff2011,Leydesdorff2012,Rotolo2016a} provided us with qualitative insights into the innovation dynamics underlying these medical contexts. We begin by providing some background on these case-studies. 

First, HPV is a sexually transmitted virus. HPV infections are relatively diffused in the population and affect several parts of the human body (e.g.\ genitals, mouth, and throat). In the 1980s, Harald zur Hausen, a virologist, and his team of researchers at the German Cancer Research Centre provided evidence of the strong association between HPV (subsequently identified as the specific HPV types 16 and 18) and cervical cancer \citep{zurHausen1976,zurHausen1987}. Zur Hausen was awarded the 2008 Nobel Prize in Physiology or Medicine for this discovery. The discovery enabled the emergence of a novel class of diagnostics for cervical cancer based on the detection of HPV DNA fragments \citep{Hogarth2012} as well as two different HPV vaccines, namely Gardasil and Cervarix, that have been commercialized by Merck and GlaxoSmithKline since 2006 and 2007, respectively. More recently, researchers have also discovered that HPV variants are also linked to head and throat cancers, producing a paradigm shift in what was already considered a mature research area \citep{Scudellari2013}. 

Second, RNAi refers to the intracellular process by which microRNA (miRNA) or small interference RNA (siRNA) are able to block fundamental gene expression pathways. Andrew Fire and Craig C.\ Mello provided evidence of the gene-silencing effect of miRNA in 1998 \citep{Fire1998}. This discovery constituted the basis for their 2006 Nobel Prize in Physiology or Medicine. The additional role of siRNA was discovered by \cite{Hamilton1999}. The potential of RNAi as a biological tool has been investigated, as the silencing of certain genes by means of RNAi could possibly be used to stop progression of diseases (e.g.\ cancers, genetic diseases, and infection agents). Given the broad range of potential therapeutic applications, the discoveries of miRNA and siRNA have increasingly attracted the attention of researchers as well as generated high expectations \citep{Sung2006, Haussecker2015}. More recently, however, the clinical development of RNAi has slowed down because of problems in translating from \textit{in vitro} to \textit{in vivo} experiments \citep{Haussecker2012,Lundin2011}.

Finally, MRI represents technological capability that has been repeatedly refined, evolving rapidly in the early 1970s from nuclear magnetic resonance (NMR) theory into a non-invasive imaging technology \citep{Lauterbur1973}. Paul C.\ Lauterbur and Peter Mansfield shared the 1993 Nobel Prize in Physiology or Medicine for this application \citep{Blume1992}. The subsequent applications of this technology within the biomedical domain alone are too many to adequately list.

\setlength{\tabcolsep}{10pt}
\renewcommand{\arraystretch}{1}
\begin{table}\footnotesize
	\caption{\label{tab:query}Search queries used to retrieve MEDLINE/PubMed data for the three examined case-studies of medical areas. $A_q$ is the number of publications returned by a given search query $q$; $V_q$ is the total number of distinct MeSH descriptors (vocabulary size) assigned to publications in the given dataset; $\langle M_q \rangle$ is the average number of MeSH descriptors per publication; $\langle \alpha \rangle$, $\sigma_\alpha$, and $Med(\alpha)$ are the average, standard deviation, and median number of MeSH descriptors in each branch per publication, respectively.}
	\centering
{\begin{tabular}{p{1.5cm}p{4cm}p{4cm}p{4cm}}
\hline\hline
&	\multicolumn{3}{c}{\textbf{Case-studies}}\\
\hline
&	 \multicolumn{1}{c}{\textbf{HPV}}& 		\multicolumn{1}{c}{\textbf{RNAi}}& 			\multicolumn{1}{c}{\textbf{MRI}}\\
\hline

$q$&	HPV*[Title] or ``Human Papilloma Virus*''[Title] or ``Human Papillomavirus*''[Title]&	``miRNA''[Title] or ``microRNA''[Title] or siRNA[Title] or RNAi[Title] or ``RNA interference''[Title] or ``interference RNA''[Title]&		``magnetic resonance imaging''[Title] or mri[Title]\\
Period&				1963-2013&	1998-2013&	1978-2013\\
$A_q$&				18,696&		17,083&		62,842\\
$V_q$&				5,777&		8,187&		11,655\\
$\langle M_q \rangle$&	13.0&		11.8&		11.0\\
$\langle C \rangle \pm \sigma_C$&	$2.5 \pm 1.7$&		$0.9 \pm 1.1$&		$1.7 \pm 1.6$\\
$\langle D \rangle \pm \sigma_D$&	$2.1 \pm 2.4$&		$3.6 \pm 2.3$&		$0.8 \pm 1.6$\\
$\langle E \rangle \pm \sigma_E$&	$2.2 \pm 1.9$&		$1.7 \pm 1.7$&		$3.3 \pm 2.1$\\
$Med(C)$&	2&	0&	1\\
$Med(D)$&	1&	3&	0\\
$Med(E)$&	2&	1&	3\\

\hline\hline
\multicolumn{4}{p{15cm}}{\footnotesize \textit{Source: Search performed by authors on MEDLINE/PubMed. Data were downloaded on May 28, 2014 for HPV and RNAi and on June 5, 2014 for MRI}}
\end{tabular}
}
\end{table}

\begin{figure}
\includegraphics[width=16cm]{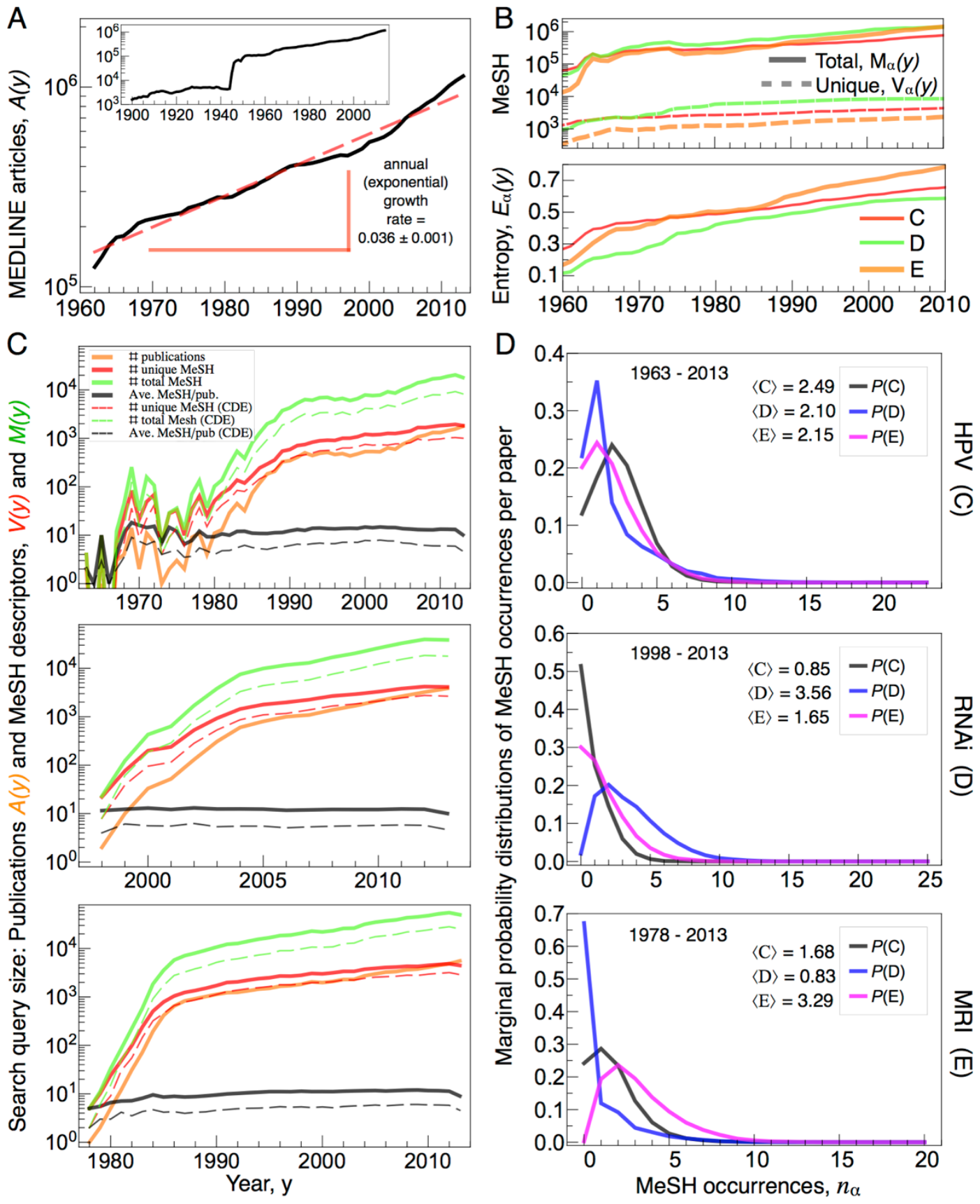}
\centering
\caption{Number of publications $A(y)$ indexed in MEDLINE/PubMed over time (A). The total number of MeSH $M_{\alpha}(y)$ and the vocabulary size $V_{\alpha}(y)$, by branch $\alpha$ and by year $y$ (B-top). The normalized entropy (efficiency) $E_{\alpha}(y)$ of the distribution of individual MeSH counts $C_{\alpha}(r,y)$ within branch $\alpha$ and by year $y$ (B-bottom). Number of publications $A_q(y)$ and associated MeSH descriptors, total $M_q(y)$ and unique $V_q(y)$, for each search query $q$ listed in \Tabref{query}; the black curves indicate that the average number of MeSH descriptors per publications (C). Empirical probability distributions, $P(n_{p,\alpha})$, representing the likelihood of observing $n_{p,\alpha}$ MeSH descriptors of type $C$, $D$, or $E$ per publication, $p$ (D). \newline\textit{Source: Authors' elaboration on the basis of MEDLINE/PubMed data.}}
\label{fig:trends}
\end{figure}

To identify publications related to these medical areas, we queried the MEDLINE/PubMed database in May/June 2014. \Figref{trends}A shows the growth of this large open medical publications index, which exhibits a 3.6\% annual growth rate since the 1960s, comprising more than 21 million article records through 2010. In addition to the growth in the MEDLINE/PubMed database size, \Figref{trends}B also shows the growth in the $C$, $D$ and $E$ branch vocabulary size, and the marked growth in the diversity (measured as the normalized entropy) of individual MeSH descriptors within each of these branches. 

In order to perform medical-area specific analyses, we used keywords-based search strategies denoted by $q$. We report the queries used in the retrieval in \Tabref{query}. For each area, we retrieve the $A_q$ publications and tally the uses of the individual MeSH descriptors $m_i$ across the vocabulary set of size $V_q$; records with no MeSH descriptors were excluded from the data.\footnote{The year 2014 was excluded since the records for this year were not yet complete at the date of this research. The majority of the publications with no MeSH descriptors are from recent years (after 2010), as they have yet to be annotated. The percent of publications with no MeSH descriptors, before the year 2011, is 2.4\% (362 publications), 3.7\% (340 publications), and 2.7\% (1,379 publications) for HPV, RNAi, and MRI, respectively. However, our results are not sensitive to MeSH annotation vacancy nor time lags in the collection of data since our quantitative measures are calculated within each year group.} HPV and RNAi are represented in the set by a similar number of publications (18,696 and 17,083 respectively), while the sample of publications associated with MRI is about three times larger (62,842 publications). Note that the collected publications span different time periods: HPV publications are represented since 1963, RNAi publications since 1998, and MRI publications since 1978. 

At the article level, each publication $p$ is characterized by an informative 'fingerprint' comprising the $N_p$ unique MeSH descriptors ($m_i$) assigned to it (with $N_p > 0$). This set of distinct descriptors form a group $\{m_1, m_2, ..., m_{N_p}\}$, which summarizes the topical content of $p$ at different levels of specificity. On the basis of this information, we calculate the average number of $m_i$ per publications, $\langle M_q \rangle$, which ranges from 11 to 13 across the examined case-studies; additional descriptive statistics are reported in \Tabref{query} and depicted in \Figref{trends}. 

The relational knowledge order has temporal features that are also important to identify, measure, and interpret. \Figref{trends}C (left) shows aggregate measures of growth by year $y$, both in terms of the number of publications $A_q(y)$, as well as the size of the MeSH vocabulary $V_q(y)$. One factor contributing to the growth trends is the increasing journal coverage by MEDLINE/PubMed over time (\Figref{trends}A). Nevertheless, we find that the trends exhibit common patterns independent of the time period: relatively fast growth during the first decade, which then saturates at approximately $10^3$ publications per year. The saturation in the growth of each research area is indicated by a kink in each count trajectory when plotted on a logarithmic scale. Compared to the early 'seed phase' growth regime, the eventually slowed growth rate likely reflects the upper limits to the research activity according to fundamental funding and labor limitations \citep{DeSollaPrice1963,Mutschke2011}.\footnote{It is also worth mentioning that MEDLINE/PubMed query service, which is based on a non-autonomous category scheme that requires human input and regulation, appears to be functioning efficiently as a search and retrieval service. This is evidenced by the saturation in the $A_q(y)$ observed for MRI. The saturation around approximately $10^3$ publications per year reflects the intended outcome of the query, which is aimed at obtaining a refined subset of research articles that are specifically related to MRI in context, without being overwhelmed by the numerous publications that use the technology in a peripheral way. Consider that in the broadest sense, the total number of publications per year \textit{loosely related or relying on} MRI is most certainly much more than $10^3$ per year (e.g., the \textit{Journal of Magnetic Resonance Imaging} publishes around $10^3$ articles per year alone).}

As discussed, the focus of our analysis is on MeSH descriptors within $\alpha = \{C, D, E\}$, with  roughly half of the MeSH descriptors we analyzed being classified within these branches.\footnote{The MeSH tree is however not a 'real' tree because there are instances of local loops within the MeSH network, with some MeSH descriptors belong to multiple $\alpha$ and/or multiple tree levels. For example, 6.9\% of the $m_i$ in the $C$, $D$, and $E$ branches have more than one branch affiliation.} The numbers of assigned MeSH descriptors, $n_{p,\alpha}$, the average, $\langle \alpha \rangle$, the standard deviation, $\sigma_\alpha$, and the median, $Med(\alpha)$ of the number of descriptors per publication for the selected branches $\alpha$ are reported in \Tabref{query} and \Figref{trends}. These support our assumption that RNAi is mostly represented by MeSH descriptors within branch $D$, while MRI is represented by descriptors within branch $E$. In contrast, HPV appears at first to be equally represented by $C$, $D$, and $E$ branches. However, the signed-rank test (Wilcoxon test) indicates that the differences in the medians between the $C$, $D$, and $E$ distributions are significantly different from zero ($p<0.001$ for each pairwise comparison). Moreover, \Figref{trends}D, which reports the distributions $P(n_{p,\alpha})$ measuring the likelihood of finding $n_{p,\alpha}$ MeSH descriptors from branch $\alpha$ in a given publication $p$, shows that in the bulk of the distribution $P(n_{p,c})$ for HPV is shifted more towards larger $n_{p,c}$ values. This further confirms that HPV is foremost situated within branch $C$.

\section{Empirical approach and results}

\subsection{Medical areas as vocabulary spaces of MeSH descriptors}
The MeSH vocabulary space is a growing system, characterized by the entry of individual $m_i$, which due to the nature of its relational construction, can be expected to affect the organization of a large subnetwork of the vocabulary space. The persistent expansion of the MeSH communication system, visualized in \Figref{trends}B, shows the total number of MeSH descriptor uses, $M_\alpha(y)$, and the vocabulary size, $V_\alpha(y)$, of all $m_i$ within a given branch $\alpha$, calculated using all of the 21 million PubMed/MEDLINE articles indexed in year $y$ (as indicated in \Figref{trends}A). Furthermore, the usage $C(m_i,y)$ of individual MeSH $m_i$, measured by the number of articles annotated by $m_i$ in year $y$, provides insight into the diversity of the MeSH descriptors. To quantify the usage diversity across all $m_i$ within a given $\alpha$, we calculated the normalized entropy (termed efficiency):
\begin{equation*}
E_\alpha(y) = - \sum_{i \in V_\alpha(y)}{\frac{P_\alpha(m_i,y)logP_\alpha(m_i,y)}{logV_\alpha(y)}}
\end{equation*}

\noindent where $P_\alpha(m_i,y) = C(m_i,y)/M_\alpha(y)$ is the frequency of $m_i$ in $y$. \Figref{trends}B shows that each 'efficiency trajectory' $E_\alpha(y)$ is increasing over time, especially in the case of branch $E$, indicative of the increasingly diverse role of 'analytic, diagnostic, and therapeutic techniques and equipment' in the medical enterprise. On the other hand, the entropy $E_D(y)$ appears to be saturating since the year 2000, suggesting a bottleneck in the discovery and developmental process of drugs and chemicals.

Thus, owing to its wide range and increasing depth and specificity, the MeSH descriptors provide a quantifiable framework for observing and measuring the co-evolutionary dynamics of innovation. While the synergetic relations across the $C$, $D$, and $E$ branches are the main focus of our analysis, in order to better understand the quantitative features of this communication system, we start here by analyzing the statistical patterns in $A_q(y)$ and $V_q(y)$ that are common across each $q$. We then use the knowledge of the underlying MeSH vocabulary statistics to visualize the usage trajectories of individual $m_i$ as well as the association patterns between the $C$, $D$, and $E$ branches. Together, these data-driven explorations provide insight and intuition for the final section, where our triple helix \textit{supply-demand-technological capabilities} model of medical innovation is specified using the mutual information in $C$, $D$, and $E$ as the measure of potential synergy.

We initiate the micro-level analysis from an intuitive starting point based on the fundamental question: How do the statistical patterns in the MeSH classification (communication) system compare with those often present in other language systems? Quantitative models of language are based on principles of 'efficient communication' (between the messengers and receivers, so as to reduce the effort of articulation while also maintaining manageable levels of noise in the communication signal) and 'organization' (according to topics, categories, and grammar) that play key roles within a vocabulary space. Zipf's law, Heaps' law, and other emergent features such as category schemes, can be used as empirical benchmarks to compare language models \citep[e.g.][]{FerreriCancho2003, Puglisi2008, Serrano2009}. In this spirit, for each medical research area $q$, we first provide evidence of statistical patterns in the vocabulary space, which indicate that the MeSH categories can be considered as statistically stable and efficient means of communication.
 
\begin{figure}
\includegraphics[width=16cm]{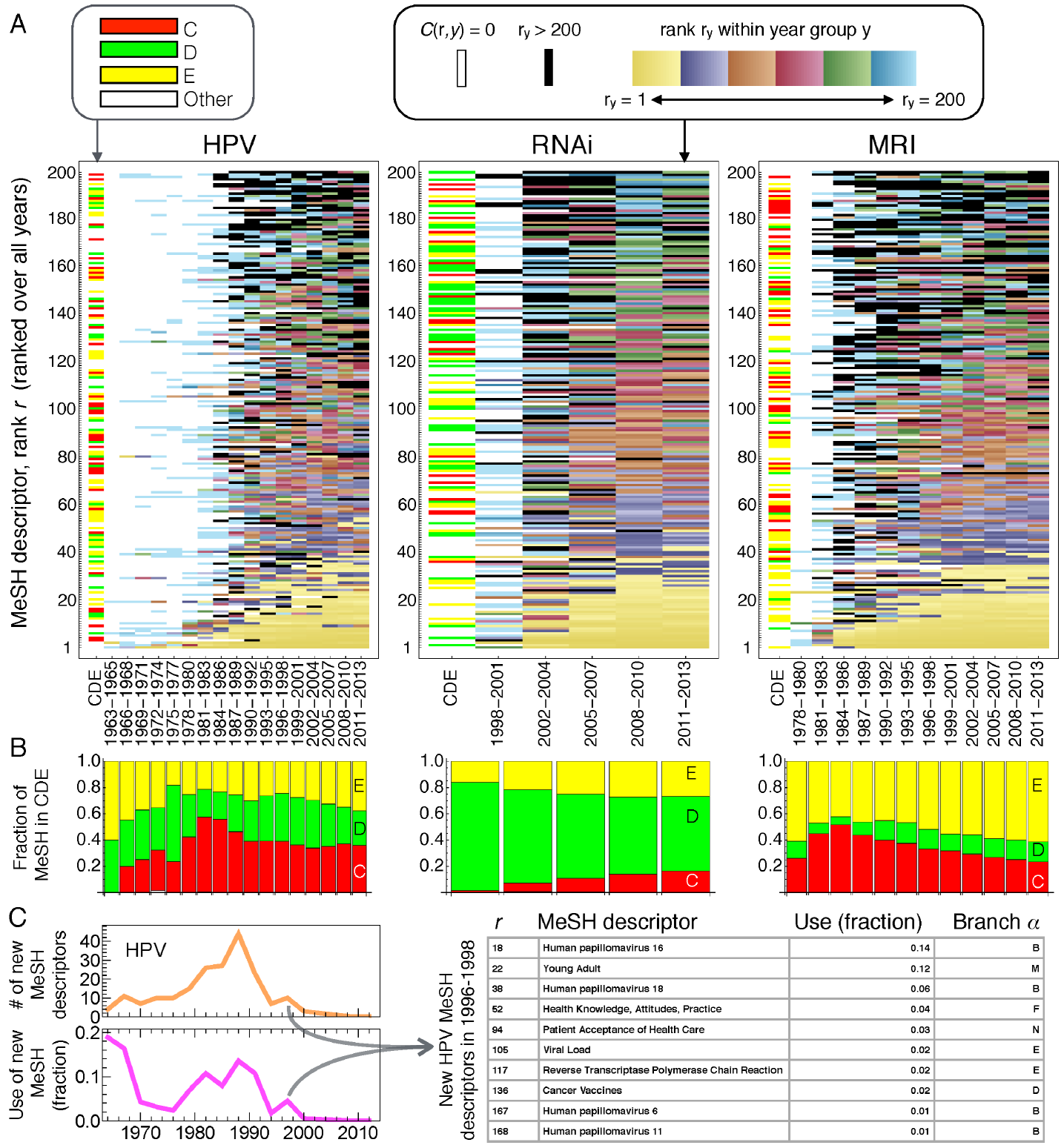}
\centering
\caption{Evolution of the selected medical areas and associated C-D-E MeSH vocabulary. (A) Evolution of the relative ranking of the top-200 most frequently used MeSH descriptors in each research area, ranked according to $C(r)$ calculated over all years. In each subpanel, the left-most column indicates the primary branch $\alpha(r)$ corresponding to each descriptor $r$; a white label indicates that the descriptor is from a main branch different than $C$, $D$, and $E$. The remaining columns indicate $r_y$, which is the local rank of the descriptor $r$ within the year group $y$. White color values indicate no observed counts for $r$ in $y$, $C(r,y) =0$; Black color values indicate that the MeSH descriptor was not ranked in the top-200 descriptors within a given $y$.  Color transitions along rows indicate the relative rise and fall of individual MeSH descriptor use within the vocabulary space. Non-sequential color mixing across columns indicates the level of rank instability within a given $y$, possible indicating a reorganization of the triple-helix configuration. (B) The fraction of the subset of MeSH descriptors in $C$, $D$, or $E$, occurring in period $y$ by branch. (C) New descriptors (indicated by white to non-white transitions along the rows in panel A) trace the evolution of each research area. The number of new descriptors can show nonlinear trends, as in the case of HPV. The number and long-term use (impact) of the new descriptors are useful indicators to identify important historical contexts, e.g.\ the identification of the cancer-related HPV variants 16 and 18 and the subsequent development of cancer vaccines, in addition to other socio-technical contexts, during the period 1996-1998. \newline\textit{Source: Authors' elaboration on the basis of MEDLINE/PubMed data.}}
\label{fig:evolution}
\end{figure}

\begin{figure}
\includegraphics[width=12cm]{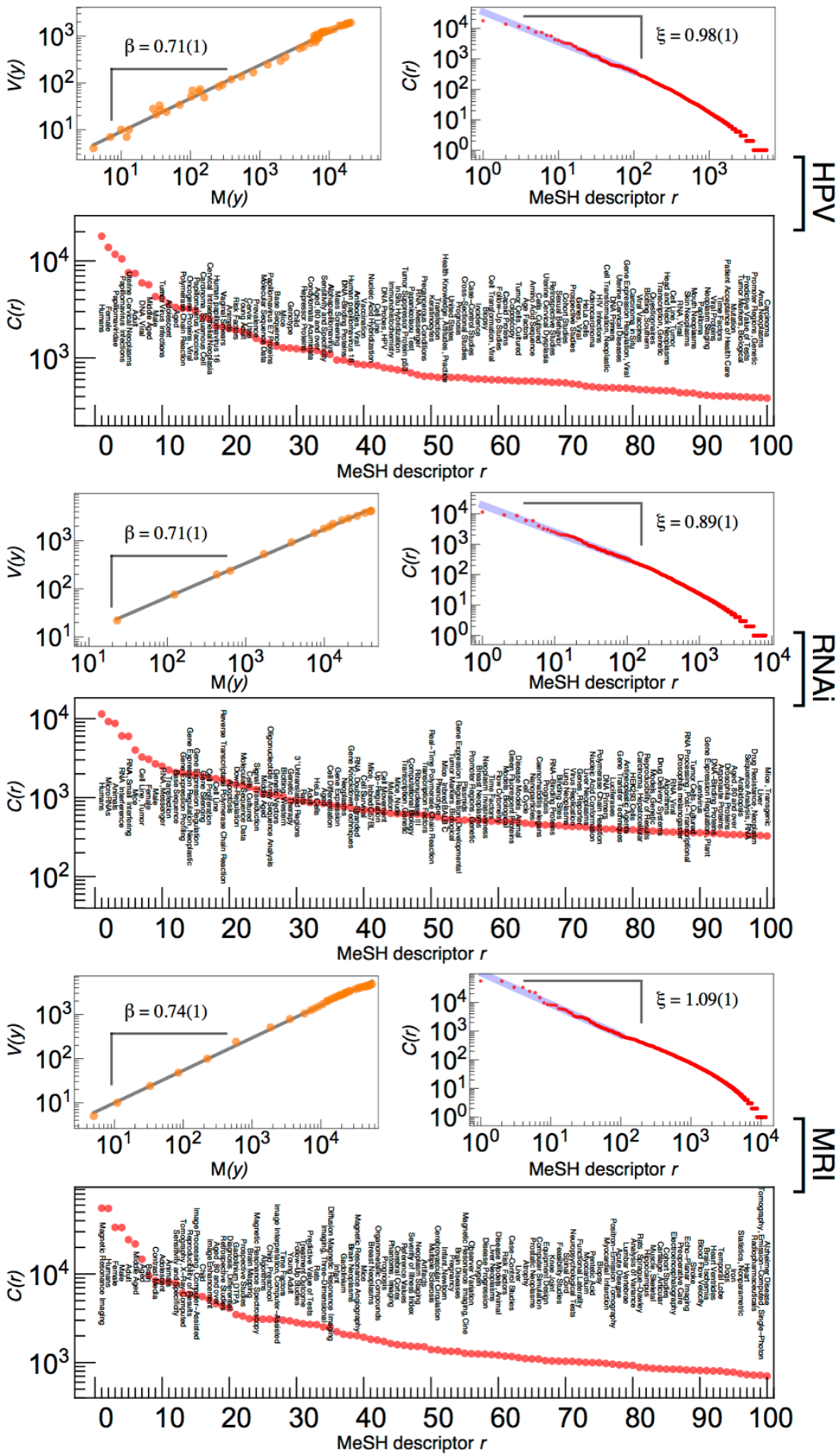}
\centering
\caption{Statistical regularities in the MeSH vocabulary space. Rank distributions of $C(r)$, the total number of papers including MeSH descriptor $r$ calculated over all years. For each research area, we demonstrate that the corpus of MeSH descriptors follows Heaps' law --- quantifying the allometric scaling of the vocabulary via the scaling exponent $\beta$ --- and Zipf's law --- quantifying the size distribution of the MeSH descriptor use via the scaling exponent $\xi$. Best-fit Heaps' and Zipf's law exponents are estimated using OLS; the standard error in the last digit shown in the parenthesis. We show the top-100 used MeSH descriptors for each $q$ to illustrate the diverse topicality of the descriptors. \newline\textit{Source: Authors' elaboration on the basis of MEDLINE/PubMed data.}}
\label{fig:regularities}
\end{figure}

We first identify the set of unique $m_i$ for each query ($q$) as the vocabulary $V_q$. We then index each $m_i$ by its rank $r$ within the respective vocabulary by sorting the MeSH descriptors in descending order according to the total number $C(r)$ of appearances in each specific $q$, i.e.\ $C(r=1) \geq C(2) \geq C(3) \geq C(r=V_q)$. \Figref{evolution} illustrates the dynamic features of the MeSH space, highlighting the entry and rank-instability among the top-200 most occurring MeSH descriptors up to 2013. Each $m_i$ is identified in the left-most column by its primary branch: $C$ (red), $D$ (green), $E$ (yellow) and 'Other' (white). Also, the identity of the top-100 $m_i$ are shown in \Figref{regularities} for visual inspection.\footnote{It is worth noting that the top few descriptors in each query are general, if not entirely contained, within the definition of each $q$. The descriptors that follow tend to become increasingly specific corresponding to lower branches of the MeSH tree (see \Figref{evolution}). Because $C(r)$ takes the form of an extremely right-skewed power-law distribution, one which is nearly scale-free except for the slight truncation for large $r$ (due to finite-size constraints because the vocabulary space is not infinite in size), there is no natural cutoff $r_c$ that separates the core $m_i$ from the rest of the $m_i$ in the vocabulary set. This is important for our analysis, as it indicates that we should not discard any $m_i$ above or below some threshold, because a natural threshold cannot be clearly identified.}
Moreover, in order to analyze the temporal information, we also tracked $C(r,y)$, the number of publications from year $y$ annotated by the MeSH descriptor $r$, and assigned a local year-specific rank $r_y$ to that $m_i$. Focusing on the top-200 $m_i$, we separated this group into six subsets (sextiles) of 33 for each $y$ based upon $r_y$. Because $r_y$ corresponds to a percentile within the top-200, we use a partitioned color scheme in \Figref{evolution}A that facilitates the visualization of the rank (in)stability of $m_i$ over time. This method shows how the rank ordering is sensitive to the entry of new $m_i$ as well as the relative growth and decay in the use of established $m_i$.

We emphasize the MeSH entry process using white cells to indicate the absence of any publications in $y$ with MeSH descriptor $r$, $C(r,y)=0$, and black cells to indicate the $m_i$ with rank $r_y > 200$. Hence, a transition from white to non-white indicates the birth of a MeSH descriptor into the corpus $V_q$, representing a growth mechanism that has notable implications on the organization and dynamics of the communication system \citep{Petersen2012b}. This can be appreciated by calculating the entry rate of new MeSH descriptors and their long-term impact. To measure the relative importance of the new $m_i$, we counted the total number of appearances of the new MeSH descriptors from its birth year through 2013 (analogous to a citation count), and divided this ``net impact'' by the total number of appearances of the top-200 MeSH descriptors over the same period. We implement this normalization in order, to control for the growth of $A_q(y)$. As an illustrative example, \Figref{evolution}C highlights the burst of MeSH entries in the period 1996-1998 for HPV. During this period, 10 new MeSH descriptors account for a 5\% share relative to the top-200 MeSH descriptors in the period following 1996, marking a noteworthy entry period within the knowledge space of HPV. 

This set of new HPV MeSH descriptors merits closer inspection. \Figref{evolution}C lists them individually, along with their associated main branch, and their individual impact, calculated as the fraction of publications after their inception in which they appeared. These new descriptors are associated with key discoveries that link certain types of cancers to specific HPV variants represented by the four $m_i$ corresponding to Human papillomavirus 6, 11, 16, and 18. Indeed, HPV 16 and 18 (appearing in 14\% and 6\% of subsequent publications) have since been established as the culprits associated with cervical cancer. The relational information contained in these MeSH entry patterns further sheds light on the complex cross-cutting nature of the combinatorial creativity in innovation processes. This is demonstrated by the fact that the new MeSH descriptors are representatives of six branches, such as ``Patient Acceptance of Health Care'' which is from the Health Care branch ($N$), with only three from the $D$ and $E$ branches.  

Furthermore, by projecting MEDLINE/PubMed articles onto the C-D-E representation, we are also able to provide quantitative insights into the historical evolution of MeSH descriptor combinations and the innovations they represent. For example, \Tabref{mesh_list} demonstrates the key relations between ``DNA, viral'' and the new methods associated with MeSH from branch $E$ (e.g.\ ``Polymerase Chain Reaction'', ``In Situ Hypbridization'') in HPV research. In the RNAi domain, the impact of ``Transfection'' methods is appreciated by noting its impact on various fundamental lines of cancer research. Similarly, the development of ``Contrast Media'' chemicals has also had a wide-reaching effect in MRI research, mediating the application of this technology to the microscopic cellular domain. As such, analyzing MeSH combinations that span all three branches may serve as a useful research evaluation indicator of novelty representing the integration of the three innovation channels.

\setlength{\tabcolsep}{10pt}
\renewcommand{\arraystretch}{1}
\begin{table}\footnotesize
	\caption{\label{tab:mesh_list}Most frequent MeSH-descriptor pairs. Given that HPV publications are mostly associated with $C$, RNAi with $D$, and MRI with $E$, the dyads that belong to the complementary branches $D$ and $E$, $C$ and $E$, to $C$ and $D$ are reported, respectively.}
	\centering
{\begin{tabular}{llc}
\hline\hline
\multicolumn{3}{c}{\textbf{HPV: 1996-2002}}\\
\textbf{Branch D descriptor}            & \textbf{Branch E descriptor}           & \textbf{Publications}\\
\hline
DNA, Viral                     & Polymerase Chain Reaction     & 523          \\
DNA, Viral                     & Vaginal Smears                & 165          \\
DNA, Viral                     & In Situ Hybridization         & 140          \\
DNA, Viral                     & Sensitivity and Specificity   & 127          \\
DNA, Viral                     & Risk factors                  & 119          \\
Oncogene Proteins, Viral       & Transfection                  & 100          \\
DNA Primers                    & Polymerase Chain Reaction     & 98           \\
DNA, Viral                     & Prevalence                    & 83           \\
DNA, Viral                     & Papanicolaou Test             & 73           \\
Oncogene Proteins, Viral       & Polymerase Chain Reaction     & 71           \\
                               &                               &              \\
\hline
\multicolumn{3}{c}{\textbf{RNAi: 2005-2008}}\\
\textbf{Branch C descriptor}            & \textbf{Branch E descriptor}           & \textbf{Publications} \\
\hline
Neoplasms                      & Genetic Therapy               & 37           \\
Liver Neoplasms                & Transfection                  & 32           \\
Carcinoma, Hepatocellular      & Transfection                  & 27           \\
Breast Neoplasms               & Transfection                  & 20           \\
Disease Models, Animal         & Genetic Therapy               & 18           \\
Neoplasm Invasiveness          & Transfection                  & 18           \\
Neovascularization, Pathologic & Transfection                  & 15           \\
Carcinoma, Squamous Cell       & Transfection                  & 14           \\
Prostatic Neoplasms            & Transfection                  & 14           \\
Neoplasm Invasiveness          & Reverse Transcriptase Polymerase Chain Reaction & 13           \\
                               &                               &              \\
\hline
\multicolumn{3}{c}{\textbf{MRI: 2002-2008}}\\
\textbf{Branch C descriptor}            & \textbf{Branch D descriptor}           & \textbf{Publications}\\
\hline
Breast Neoplasms               & Contrast Media                & 169          \\
Liver Neoplasms                & Contrast Media                & 140          \\
Myocardial Infarction          & Contrast Media                & 114          \\
Brain Neoplasms                & Contrast Media                & 90           \\
Neovascularization, Pathologic & Contrast Media                & 76           \\
Disease Models, Animal         & Contrast Media                & 76           \\
Breast Neoplasms               & Gadolinium DTPA               & 68           \\
Myocardial Infarction          & Gadolinium DTPA               & 63           \\
Carcinoma, Hepatocellular      & Contrast Media                & 60           \\
Neoplasms                      & Contrast Media                & 55    \\

\hline\hline
\multicolumn{3}{p{15cm}}{\footnotesize \textit{Source: Source: Authors' elaboration}}
\end{tabular}
}
\end{table}

Returning to the entry-exit process illustrated in \Figref{evolution}A, a transition between black and non-black cells indicates notable rank transitions out of the set of top-200 used MeSH descriptors. Interestingly, the black cells are dispersed across the entire range, consistent with burstiness in topicality \citep{Chen2006}, reflects how the research front captured by scientific literature can quickly change direction, either by chance or by external redirection (e.g.\ by funding initiatives). Nevertheless, the top-33 MeSH descriptors in each query (yellow scale), are relatively stable and represent the core vocabulary subset. The rest of the MeSH descriptors exhibit variable rank mixing. As it will be discussed in the next section, this variability is consistent with the innovative dynamics represented by the coupling of the $C$, $D$, and $E$ branches of the MeSH classification. It should also be noted that the level of rank-stability provides valuable information concerning the organization and dynamics of complex systems \citep{Blumm2012}. 

Insight into the evolution of the knowledge order can be obtained by applying a system-size scaling analysis, as commonly implemented in computational linguistics. Consider the two basic quantities that define each $q$: (i) the number of distinct MeSH descriptors (the size of the vocabulary) used in a given year, $V_q(y)$, and (ii) the total number of MeSH descriptors used in that same year, $M_q(y)$, which is roughly proportional to the total number of papers $A_q(y)$ (see \Figref{trends}). Heaps' (\citeyear{Heaps1978}) law, $V = b M^\beta$, represents a simple yet revealing allometric scaling relation between a corpus size measure, $M$, and a diversity measure, $V$. The scaling exponent $\beta$ quantifies how the topical diversity (variety) of a research area --- measured by the vocabulary size $V_q(y)$ --- depends on the conceptual volume of the research area, $M_q(y)$. Furthermore, since $M_q(y)$ and $V_q(y)$ tend to grow over time, the value of the scaling exponent $\beta$ provides indication of the marginal returns to $M_q(y)$ that may arise with the inclusion of a new descriptors: the derivative $dM/dV = b^{-1/\beta} V^{(1/\beta)-1}$ is increasing with $V$ for $\beta<1$ and decreasing with $V$ for $\beta>1$. (The constant $b$ is a mathematical artifact, and is not important here since the allometric relation is fundamentally scale free, i.e.\ $\beta$ does not depend on the units chosen to measure $M$ or $V$.) 

The value $\beta<1$ is commonly found for systems characterized by efficiency-oriented principles of organization \citep{Bettencourt2007}. For each $q$ we observe sub-linear values, $\beta \approx 0.7$, indicating an economy of scale, whereby there is a decreasing marginal need for new MeSH descriptors, but an increasing marginal returns per new $m_i$ --- likely arising due to the tree-like network structure that is fundamental to the MeSH classification system.  Sub-linear $\beta$ values have also been previously reported for seven different languages captured in extensive written corpora \citep{Petersen2012a}. 

Heaps' law is frequently complemented by a second statistical regularity: Zipf's (\citeyear{Zipf1949}) law, which measures the size-distribution of the entities. Knowledge of the size distribution --- here the distribution of $C(r)$ --- can provide insight into the organization and relational structure of the communication system by ruling out certain underlying processes while pointing to others --- e.g.\ preferential attachment, structure-mediated growth, life-cycles, and the role and strength of communicator noise \citep{FerreriCancho2003, Newman2005, Amaral1998, Petersen2014,Petersen2012a}. The possible functional forms of $C(r)$ (e.g.\ linear, exponential, power-law, discrete generalized beta distribution) can be distinguished by plotting on a $log-log$ scale. Our results indicate that the $C(r)$ are indeed approximately distributed following Zipf's law: $C(r) = C(1) r^{-\xi}$, that is, as a power-law distribution with scaling exponent $\xi \approx 1$, for several orders of magnitude.  

The statistical regularities captured by $\beta$ and $\xi$ are robust across the three examined case-studies of medical areas, thus indicating that each of the MeSH vocabularies forms a relatively comprehensive representation of a respective knowledge space. These results legitimate a quantitative decomposition of each $q$ into its fundamental (topical) knowledge representations. As such, we expect that reasonable queries, in general, can be expected to map onto a representative knowledge order featuring these statistical regularities.

Albeit entirely descriptive, these data-mining exercises serve as formal data-exploration procedures that are necessary prerequisites in large data-centric analyses: had the statistical regularities failed to emerge, we would not have been legitimated  to continue with a comparative case-study analysis. In summary, we argue in this section that the composition and dynamics of the MeSH classification for the three case-studies of medical areas under study can be considered as interactions among three knowledge orders. This provides us with a starting point for investigating the mutual relations and potential synergy contained in the triple helix of $C-D-E$ categories. In the following sections we further investigate our central hypothesis that the evolution of each medical area is mediated by synergies among the demand-supply-technological capabilities communication channels.

\subsection{Tracing mutual information and redundancy in \textit{C--D--E}}

To evaluate the mutual information between $C$, $D$ and $E$ branches, we use the Shannon's definition of information for a single discrete random variable $x$ \citep{Shannon1948}:
\begin{equation*}
H_x = - \sum_{i}{P(x_i)log_2P(x_i)}
\end{equation*}

\noindent where the sum is over the entire range of the variable $x$. This is defined in bit units when using $log_2$. On the basis of this definition, the pairwise mutual information (transmission) between two variables $x$ and $y$ is defined as: 
\begin{equation*}
T_{xy} = H_x+H_y-H_{xy}
\end{equation*}

\noindent where the joint entropy is $H_{xy} = - \sum_{i,j}{P(x_i,y_j)log_2P(x_i,y_j)}$. $T_{xy}$ by definition can assume only positive values \citep{Theil1972}. In the case of three discrete random variables $x$, $y$, and $z$ the mutual information is instead defined as:
\begin{equation*}
T_{xyz} = H_x+H_y+H_z-H_{xy}-H_{xz}-H_{yz}+H_{xyz}
\end{equation*}

\noindent where the joint entropy is $H_{xyz} = - \sum_{i,j,k}{P(x_i,y_j,z_k)log_2P(x_i,y_j,z_k)}$. $T_{xyz}$ can assume both positive and negative values \citep{Abramson1963,McGill1954,Yeung2008}. \cite{Leydesdorff2014b} suggested that positive $T_{xyz}$ values correspond to relational integration, while negative values correspond to 'synergetic integration' in terms of correlations and positions. The reduction in the uncertainty is generated (as mutual redundancy) because the interaction terms have different meanings in each of the full sets and can thus be counted twice. In other words, a communication field or 'overlay' \citep{Etzkowitz2000} can be generated on top of the sum of communication channels because of possible next-order loops in the communication \citep{Ivanova2014}. 

The $T_{xyz}$ indicator enables us to quantify this effect \citep{Jakulin2005,Yeung2008}, with a focus on $C$, $D$ and $E$ branches. Three interacting (pairwise) information channels can be defined, i.e. $C-D$, $C-E$ and $D-E$. These can be used to assess the $T_{CD}$, $T_{CE}$, and $T_{DE}$ bilateral mutual information as well as the $T_{CDE}$ trilateral mutual information. When these are considered together, they give rise to complex features in a 'triple helix' configuration that can be additional (or subtractive) with respect to the sum of single-channel systems \citep{Krippendorff2009b}. 

In principle, every distinct triplet ($m_i$, $m_j$, $m_k$) of MeSH descriptors assigned to each publication can be interpreted as a ``knowledge encoding''. However, the combinatorial space of all distinct $m_i$ triplets corresponds to a language system with an extremely large alphabet and would be inefficient and easily dominated by fluctuations (noise). By way of example, consider the central dogma of the genetic transmission of biological information, which is based upon the 64-letter codon alphabet that is translated (with redundancy) into a code based on a 20 to 22 letter amino acid alphabet. The built-in redundancy in the translation steps of this communication system has implications for the efficiency, directionality, and error correction capability of the genetic code \citep{Yockey2005}. 

In order to avoid the problems associated with an extremely large alphabet, for theoretical and computational reasons, we explore more redundant representations of the knowledge space. Specifically, we do not track all relational combinations of MeSH descriptors ($m_i$, $m_j$, $m_k$), but instead count the number of MeSH descriptors from each of the $C$, $D$, and $E$ branches, ($n_{p,C}$, $n_{p,D}$, $n_{p,E}$), assigned to a given publication $p$. Because the number of MeSH descriptors assigned to each publication can be interpreted in various ways, we proceed with a heuristic that investigates different projections of a given publication $p$ onto $C-D-E$. All three definitions are based on a count vector, $Z_{p,CDE} \equiv (z_{p,C},z_{p,D},z_{p,E}) = f(n_{p,C},n_{p,D},n_{p,E})$, where each component $z_{p,\alpha}$ depends on $n_{p,\alpha}$ via a function $f: n_{p,\alpha} \Rightarrow z_{p,\alpha}$. 

In the first definition of $f$, we consider $z$ as a binary indicator: $z_{p,\alpha}$ is equal to $1$ if $n_{p,\alpha}>0$, i.e.\ if at least one MeSH descriptor from the $\alpha$ branch is assigned to $p$, and $0$ otherwise. In the second definition of $f$, we use the median value as threshold: $z_{p,\alpha}$ is equal to $1$ if $n_{p,\alpha}>Med(\alpha)$, and $0$ otherwise (median value are listed in \Tabref{query}). In the final definition of $f$, we use a full count definition, i.e.\ $z_{p,\alpha}=n_{p,\alpha}$. Since the first two definitions reduce the branch representation to a binary field, they significantly reduce the observation (communication) space to the eight possible combinations corresponding to $Z_{p,CDE} = (\pm1,\pm1,\pm1)$. Thus, while a binary representation may be more efficient, it is also an extremely redundant communication system. The full count representation, however, is less efficient: there are about $20^3 = 8000$ possible ``encodings'' because the maximum $n_{p,\alpha}$ value we observed was roughly 20 for each of the three $C$, $D$, $E$ categories.

\subsubsection{Defining confidence intervals}
In the analysis that follows, we shall first investigate the conditions under which the bilateral ($T_{CD}$, $T_{CE}$, and $T_{DE}$) and trilateral ($T_{CDE}$) mutual information are significantly different from zero.\footnote{ It is worth noting that mutual information is equal to zero when the underlying random variables are independent, i.e.\ the joint distribution of the random variables is equal to the product of the individual distributions (e.g.\ $P(n_C,n_D)$ is equal to $P(n_C)P(n_D)$).}  In order to establish statistical significance levels, we next outline a shuffling null model that enables us to specify confidence intervals for the empirically observed $T_{\alpha_1\alpha_2}$ and $T_{CDE}$ values. 

Our shuffling method operates at the level of publications, which are each characterized by $n_{p,\alpha}$ descriptors. For each year, we count the total number of descriptors from all $A_q(y)$ publications in that year: 
\begin{equation*}
M(y) = \sum_{p=1}^{A(y)}{(n_{p,C} + n_{p,D} + n_{p,E})}
\end{equation*}

We then shuffle these MeSH descriptors, maintaining (i) the total number $M(y) = \sum_{p=1}^{A(y)}{n_{p,\alpha}}$ of descriptors from each branch per year, and (ii) the total number ($n_{p,C} + n_{p,D} + n_{p,E}$) of descriptors assigned to each publication. In other words, the procedure preserves the average, but not the median $n_{p,\alpha}$ since the marginal probability distributions may change. However, the numbers of publications and descriptors shown in the panels on the left of \Figref{trends} and also in \Figref{regularities} remain unchanged since our shuffling is performed only within the publications of a given $y$ and $q$. 

This shuffling method reveals how much the mutual information $T$ may depend on the specific allocation of MeSH descriptors at the publication level. Values of $T_{CD}$, $T_{CE}$, $T_{DE}$ and $T_{CDE}$ that exceed the confidence intervals established by randomization indicate that the pairwise and triple-helix relations are in excess of what one can expect to be contributed to the $C-D-E$ signal by semantic noise, which may very well exist in the process of assigning MeSH descriptors to individual publications (e.g.\ indexer effects).

\subsubsection{The pairwise mutual information}
\begin{figure}
\includegraphics[width=15cm]{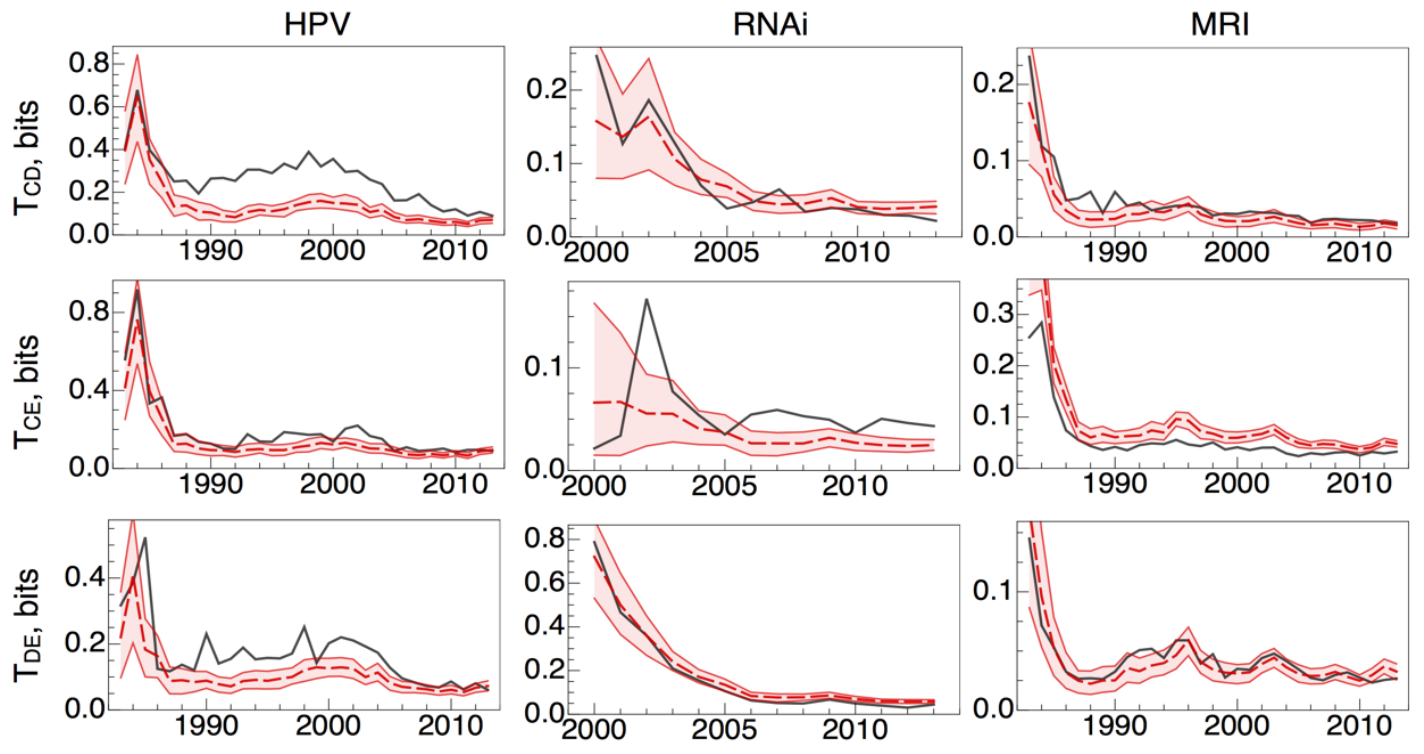}
\centering
\caption{Pairwise mutual information calculated using the full count representation $z_{p,\alpha}=n_{p,\alpha}$. In each panel, the solid black curve is $T_{\alpha_1\alpha_2}(y)$ calculated from the real data, (with $\alpha_1, \alpha_2 = \{C, D, E\}$ and $\alpha_1 \neq \alpha_2$); the red dashed curve is the mean value $\langle T_{\alpha_1\alpha_2}^{rand} \rangle$ and the red shaded area represents the empirical 90\% interval calculated from 100 randomized time series. \newline\textit{Source: Authors' elaboration on the basis of MEDLINE/PubMed data.}}
\label{fig:double}
\end{figure}

The pairwise mutual information $T_{\alpha_1\alpha_2}$ corresponding to the full count representation $z_{p,\alpha}=n_{p,\alpha}$ is depicted in \Figref{double}, where $\alpha_1, \alpha_2 = \{C, D, E\}$ and $\alpha_1 \neq \alpha_2$. As noted above, for each query and each count map definition, we shuffled the $M_q(y)$ descriptors in year $y$ and calculated $T_{\alpha_1\alpha_2}^{rand}(y)$. We constructed 100 randomized $T_{\alpha_1\alpha_2}^{rand}(y)$ time series to estimate 90\% confidence intervals corresponding to the value for which only 5\% of the randomized $T_{\alpha_1\alpha_2}^{rand}(y)$ were above or below, respectively. We also use the upper and lower confidence-interval bounds as indicators for the degree to which $T_{\alpha_1\alpha_2}(y)$ is significantly different from zero. 

For each $q$ and each $\alpha_1-\alpha_2$ pair, \Figref{double} shows a decreasing $T_{\alpha_1\alpha_2}(y)$, which saturates between $0.05$ to $0.2$ bits. This is consistent with the expectation of outward expansion of an emerging research field because of increasing specialization, corresponding to a reduction of the mutual dependence between $\alpha_1$ and $\alpha_2$. In other words, this indicates that MeSH descriptors of each $\alpha_1$ are becoming more distinctly informative (decoupled) in increasingly elaborated discourses, rather than being locked into the space of initial combinations. There are, however, short periods in which$T_{\alpha_1\alpha_2}(y)$ increases (e.g.\ for HPV around 2000 and MRI around 1996). This provides indication of increasing dyadic coupling, possibly originating from paradigm shifts that collectively reconfigure the direction of an entire field. The comparison of the magnitudes of the $T_{\alpha_1\alpha_2}(y)$ for each $q$ shows that $T_{CE}(y)$ values are on average the largest, with $T_{DC}(y)$ being slightly larger than $T_{DE}(y)$. This lends support to our conclusion that the $C-E$ channel represents the strongest dyadic link (as discussed below and depicted in \Figref{results}). 

The analysis also reveals various periods when $T_{\alpha_1\alpha_2}(y)$ in the empirical data is above the 90\% confidence interval, thus indicating an increase in the mutual information above what is expected from just the background coupling of $m_i$. \Figref{double} suggests that these periods tend to occur in each $T_{\alpha_1\alpha_2}(y)$ where the dominant branch representing each query, $\alpha_q$, is not included, i.e.\ $T_{DE}(y)$ for HPV, $T_{CE}(y)$ for RNAi, and $T_{CD}(y)$ for MRI. Because our identification strategy identifies each research area with a main branch, thereby serving as its 'entry point' into the multidimensional model, the pattern of excess $T_{\alpha_1\alpha_2}(y)$ between the adjacent branches points to presence of higher-order $C-D-E$ coupling. The instances of mutual information in excess of the 90\% confidence intervals also provide quantitative indication that the three branches are coupled via tri-lateral relations, which are 'blinded' in the analysis of bilateral relations. In other words, a communication field (or 'overlay' of communications) is indicated as significantly contributing to reducing the uncertainty that prevails in the configurations among the three branches. Investigating the sources of this excess reduction of uncertainty is an avenue for further research, likely requiring closer inspection of micro-level analysis of dyadic, triadic, and higher-order correlations between individual $m_i$.

\subsubsection{The mutual information in three dimensions}
The dynamics of the mutual information in three dimensions can provide insight into the evolution of redundancy (synergy) within the $C-D-E$ triple-helix relations. The yearly mutual information in three dimensions is defined as:
\begin{equation*}
T_{CDE}(y)=H_C(y)+H_D(y)+H_E(y)-H_{CD}(y)-H_{CE}(y)-H_{DE}(y)+H_{CDE}(y)
\end{equation*}

As discussed, $T_{CDE}$ can be negative, thereby indicating a reduction of uncertainty due to increasing redundancy. To appreciate the origin of negative values, it may be helpful to rewrite the equation above in terms of two contributions, as follows: 
\begin{equation*}
T_{CED}(y)=[T_{CD}(y)+T_{CE}(y)+T_{DE}(y)] + [H_{CDE}(y)-H_C(y)-H_D(y)-H_E(y)]
\end{equation*}

Due to the subadditivity property,  $H(x_1,x_2, ..., x_n) \leq \sum_{i=1}^{n}{H(x_i)}$, which hold for any dimension $n \geq 2$, the second bracket makes a negative contribution, whereas the terms in the first bracket are strictly positive.\footnote{In two dimensions the inequality $0 \leq \sum_{i=1}^{2}{H(x_i)-H(x_1,x_2)}=T_{12}$ corresponds to the exact definition of the mutual information, thus establishing its positivity. In three dimensions, $0 \leq \sum_{i=1}^{3}{H(x_i)-H(x_1,x_2,x_3)}=-T_{123}+\sum_{ij}^{3}{T_{ij}}$, and in four dimensions, $0 \leq \sum_{i=1}^{4}{H(x_i)-H(x_1,x_2,x_3,x_4)}=-T_{1234}+\sum_{ijk}^{4}{T_{ijk}}+\sum_{ij}^{6}{T_{ij}}$, where the sums are over the permutations of the indices. It follows (inductively) that in any given dimension $n$, we may conjecture that there exists a combination of mutual information corresponding to $0 \leq \sum_{i=1}^{n}{H(x_i)-H(x_1,x_2,..., x_n)}$, which is by definition positive (or zero in the null case of complete independence). Note that the sign of the $n^{th}$-order mutual information alternates consequentially with each additional dimension \citep{Leydesdorff2014b}.} 

In other words, negative $T_{CDE}(y)$ values arise when the differences in entropy in the second bracket outweigh the contributions from the pairwise mutual information. This scenario represents 'synergetic' resonance, generating reduction of uncertainty, among the three helices \citep{Leydesdorff2014a, Leydesdorff2014b}. As discussed, a reduction of uncertainty can also be considered as a niche that can provide opportunities for innovation \citep{Schot2007}. 

\begin{figure}
\includegraphics[width=15cm]{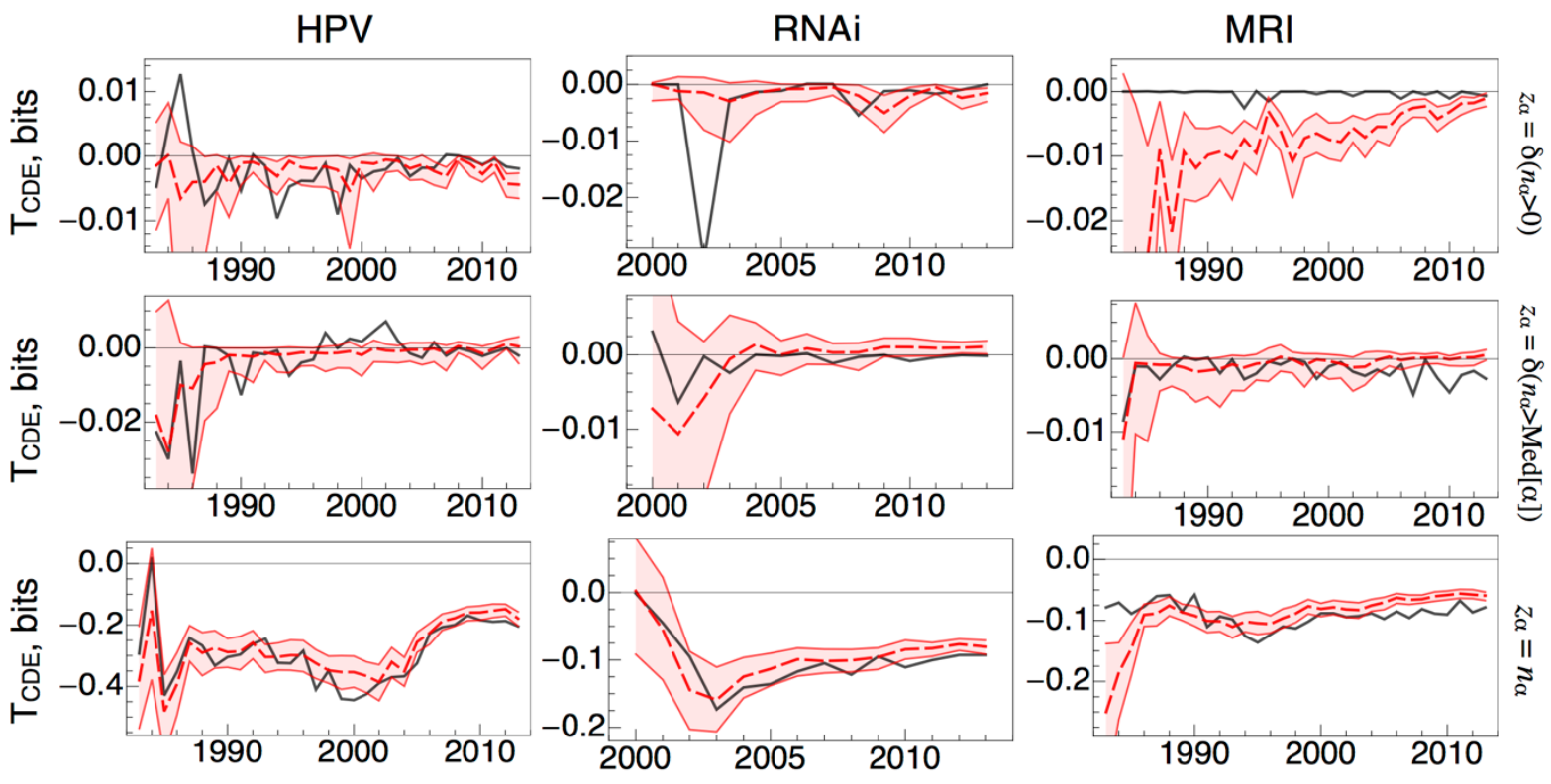}
\centering
\caption{$C-D-E$ triple-helix mutual information. Mutual information calculated using the binary count representation (top row), the median threshold method (middle row) and the full count representation (bottom row). In each panel, the solid black curve is $T_{CDE}(y)$ calculated from the real data; the red dashed curve is the mean value $\langle T_{CED}^{rand} \rangle$ and the red shaded area represents the empirical 90\% interval calculated from 100 randomized time series. \newline\textit{Source: Authors' elaboration on the basis of MEDLINE/PubMed data.}}
\label{fig:triple}
\end{figure}

\Figref{triple} shows $T_{CDE}(y)$ for each query and using each of the three MeSH count definitions. This figure reports the 90\% confidence intervals (shaded regions) calculated from 100 randomized  time series. To reemphasize, the mean random trajectory $\langle T_{CED}^{rand} \rangle$ measures the background (systemic) redundancy, since all paper-level information is eliminated by the randomization procedure. Negative deviations indicate more synergy than expected by random mixing, and positive deviations indicate less synergy than expected from random fluctuations.

The differences among the counting definitions are rather pronounced for $T_{CDE}$ (in \Figref{double}). For the first two counting methods, which apply thresholds to $n_{p,\alpha}$ thereby significantly reducing the information content of the communication channels, we observe $T_{CDE}(y) \approx 0$. Values around zero indicate that the pairwise mutual information are perfectly balanced by the excess uncertainty. As such, the synergies are not evident using these two simple counting schemes. 

When using the full-count method, we find that most $T_{CDE}(y)$ values are significantly negative, ranging from $-0.1$ to $-0.4$ bits, depending on $q$ and $y$. In other words, the nonzero $T_{CDE}(y)$ indicates that the $C-D-E$ interaction is indeed significant, further suggesting the limitations of the analysis of bilateral information in tracing innovation dynamics in the medical context. However, most of the empirical $T_{CDE}(y)$ curves are contained within the 90\% confidence interval bands, meaning that the synergy is, most of the time contained, at the systemic level. When $T_{CDE}(y)$ is not contained by the 90\% confidence interval bands, it is typically more negative, indicating synergy significantly more than expected from the background resonance of $C-D-E$. The most apparent periods of synergy occur for HPV during the period 1999-2001, for RNAi during the period 2010-2012, and for MRI from 2004 to 2013. 

Decomposition of the three pairwise mutual information trajectories for HPV in \Figref{double} reveals that $T_{CD}(y)$, owing to a relatively large $H_D(y)$, is responsible for the deviations in $T_{CDE}(y)$. This increase of uncertainty in the configuration indicates a 're-organization' in the chemical and drugs branch ($D$), consistent with the entry of ``Cancer vaccines'' and other important MeSH descriptors representing breakthroughs in the years 1996-1998 immediately preceding, as well as with the intense research efforts focused on the development of molecular biology-based diagnostic technologies \citep{Hogarth2012}, which sourced from chemicals (e.g.\ proteins) comprising $D$. 

For additional examples on the interaction between $C-D-E$, consider the new $m_i$ highlighted in \Figref{evolution}C, which capture the shifting of the research front during the subsequent decade towards the rapid development of HPV (cancer) vaccines. Moreover, the excess redundancy exhibited by MRI from 1984-present is owed to the increasing number of diseases and biomarkers that contributed to the relevance of MRI as a diagnostic instrument \citep{Blume1992,vonHippel1988}, which translated into additional redundancy in the $C-D-E$ relations. The period indicated for RNAi (2010-2013) corresponds with the current period of relative stagnation for technical reasons, which, in turn, offers opportunities for advancements and innovation in new directions \citep{Rotolo2016a}.

\section{Discussion and conclusions}
We developed a triple helix model of innovation based on three interacting sub-dynamics: \textit{supply} of and \textit{demand} for innovation, and \textit{technological capabilities} that are available to generate products, processes, and services. Such a specification was operationalized with the use of entropy statistics of mutual information among three interacting dimensions \citep{McGill1954, Yeung2008}. We use the indicator $T_{xyz}$ to measure the reduction of uncertainty and increasing redundancy in the innovation process \citep{Leydesdorff2014b}. 

Given the persistent uncertainty that is present in all the stages of the medical innovation process \citep[e.g.][]{Consoli2016,Gelijns2001}, the quantitative measurement of uncertainty using information theory is particularly apt for testing our model. In order to operationalize a study of the mutual information among \textit{supply}, \textit{demand}, and \textit{technological capabilities}, we leveraged the vast, consistent and detailed vocabulary of terms, namely MeSH descriptors ($m_i$), used to describe the topical content of individual medical publications. More specifically, we used the set of descriptors associated with the ``Diseases'' ($C$), ``Drugs and Chemicals'' ($D$), and ``Analytic, Diagnostic, and Therapeutic Techniques and Equipment'' ($E$) branches of the MeSH classification as knowledge representations of \textit{supply}, \textit{demand}, and \textit{technological capabilities}, respectively, and the co-occurrence of descriptors in publications as information for the longitudinal analyses of (bilateral and trilateral) mutual information between the dimensions discussed above. 

Due to the data-intensive nature of our analysis, we began with prerequisite testing of our measurement instruments by examining the MeSH classification from a language-communication system perspective. The prevalence of $m_i$ from the $C$, $D$, and $E$ branches provided the opportunity to identify two statistical regularities: (i) the relationship between the number of unique MeSH descriptors and the total number of descriptors used in a given year follows Heaps' law with a scaling coefficient $\beta \approx 0.7$; while (ii) the relationship between the number of times a descriptor is assigned to publications, its use $C(r)$, and the rank of this descriptor in comparison to other descriptors, $r$,  follows a Zipf's law with scaling coefficient $\xi \approx 1.0$. These statistical regularities, which were common to each $q$, point to the capacity, stability, and efficiency of the MeSH 'communication systems' representing knowledge orders. Interestingly, the sublinear ($\beta < 1$) Heaps' coefficient values point to the presence of increasing marginal returns associated with the addition of a new descriptor to the vocabulary space of a given knowledge area.

We then analyzed the descriptors composing the MeSH vocabulary of three case-studies of medical areas, namely RNAi, HPV, and MRI. We selected these case-studies as representatives of the three medical innovation dimensions, thus representing different 'entry perspectives' for our model. Our analyses confirmed this \textit{ex-ante} selection: HPV was mostly represented by descriptors within $C$ (\textit{demand}), RNAi by descriptors within $D$ (\textit{supply}), and MRI by descriptors within $E$ (\textit{technological capabilities}). Our results revealed the existence of a stable core set of descriptors that delineate a given medical area as well as of various unstable descriptors that enter in and exit from the vocabulary over time (e.g.\ those descriptors that move in and out of the top-200 descriptors rankings in \Figref{evolution}). 

To give an example of the significant cross-cutting information contained in the entry dynamics of the MeSH descriptor vocabularies, we refer to the case of HPV research, for which descriptors of cervical-cancer specific HPV variants (``Human papillomavirus 16, and 18'') entered into the vocabulary during the years 1996-1998 (see \Figref{evolution}C). During this same period, several other descriptors entered in the vocabulary of HPV research (e.g.\ ``cancer vaccines'', ``young adult females'', ``Patient Acceptance of Health Care'' and ``Health Knowledge, Attitudes, Practice''). These additional descriptors are indeed central to recent debates concerning the ethics of cancer vaccines, public anti-vaccination movements, and government vaccination policies and interventions \citep{Horne2015}.

In other words, our investigations into the dynamic properties of the MeSH categories identified both change --- as in the growth of the communication system measured by $V_q(t)$ --- and stability --- in the form of statistical regularities over time and across $q$ captured by Zipf's and Heaps' laws. Stability prevails and provides us with a baseline against which one can measure. We then developed and implemented a \textit{shuffling null model} for MeSH descriptors that enabled us (i) to assess the extent to which values of mutual information fall within an expected or unexpected range over time, (ii) to identify the significance of the non-zero mutual information values, and (iii) to detect the possible origin of the deviations from the empirical confidence intervals (e.g.\ MeSH indexer effects).

\begin{figure}
\includegraphics[width=\linewidth]{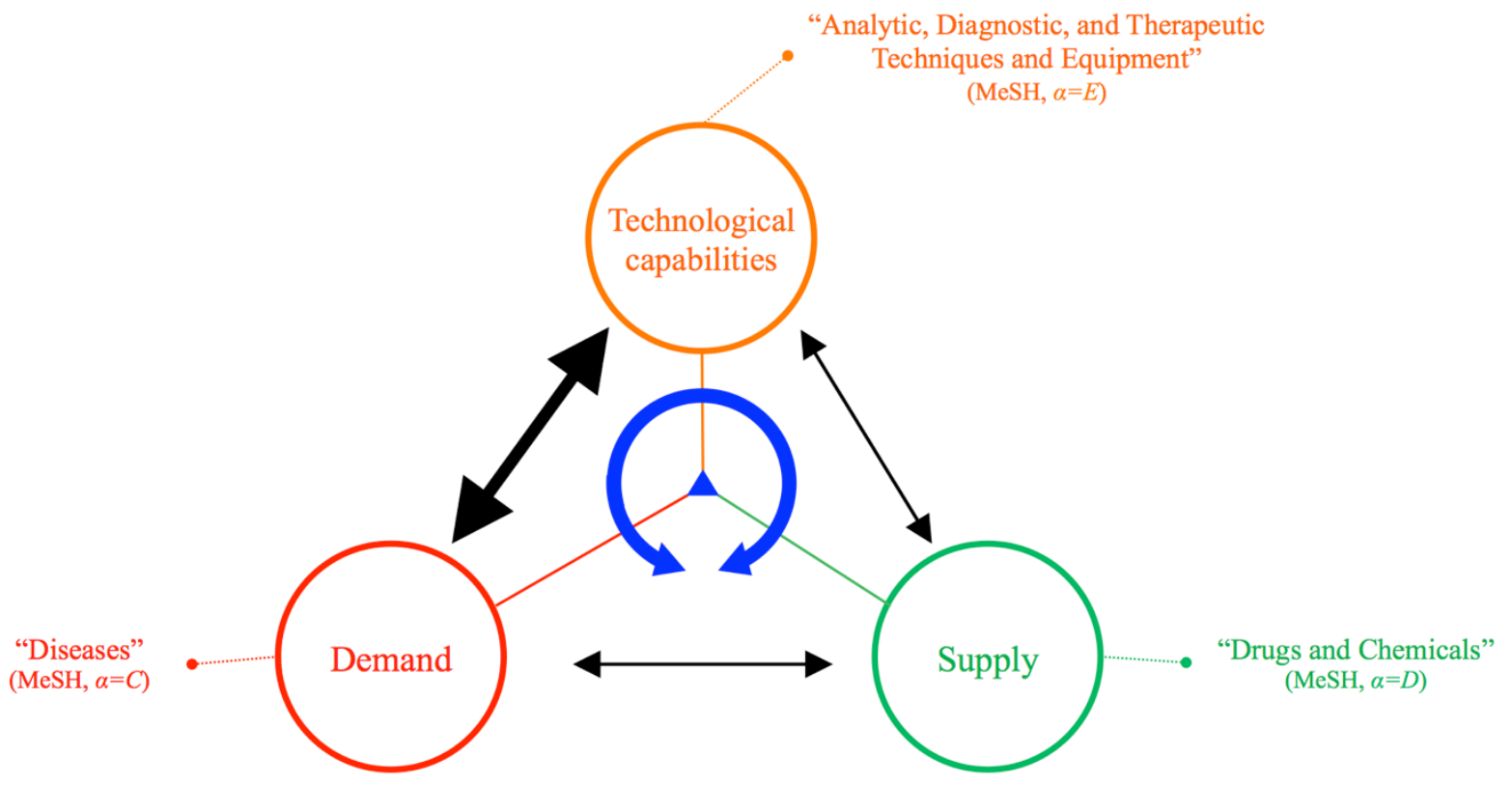}
\centering
\caption{Stylized empirical configuration of feedback loops based on interactions among \textit{supply}, \textit{demand}, and \textit{technological capabilities}. Arrow thicknesses represent the strength level determined in our quantitative analysis, with $C-E$ being the strongest, and the $C-D-E$ interaction represented by the loop also being identified as an inextricable component of the medical innovation model. Our model does not provide any information on the relative interaction strength in opposing direction, hence the arrows lack directionality, as compared to \Figref{model}. \newline\textit{Source: Authors' elaboration.}}
\label{fig:results}
\end{figure}

To highlight the crucial role of \textit{technological capabilities} as a driving force of the medical innovation process, it is worth noting that the percentage of MeSH descriptors in the $E$ branch has increased during the entire period following the mid-1990s for each case-study (see \Figref{evolution}B). Furthermore, cross-comparison of the probability distributions $P(n_{p,\alpha}$) in \Figref{trends} and the curves in \Figref{double} indicate that $C-E$ (\textit{demand-technological capabilities}) relation is the strongest of the three bilateral channels. Indeed, the growth in the prevalence of $m_i$ from branch $E$ is further complemented by a surge in the late 1980s in the diversity of $m_i$ from branch $E$, quantified using the entropy efficiency index $E_\alpha(y)$, which we calculated for the entire medical literature, comprising more than 21 million MEDLINE articles through 2010 (see  \Figref{trends}B). These trends likely arise due to \textit{technological capabilities} representing both inputs to and outputs of the scientific production function \citep{Stephan1996}. For example, MRI is used in biomedical research as well as in diagnostic patient care, serving as a widespread image analysis tool, from the micro scale of individual molecules (e.g.\ disease biomarkers) to the macro scale of organs (e.g.\ dynamics of the brain and heart). 

We then longitudinally examined mutual information in two (namely, $T_{CD}$, $T_{CE}$, and $T_{DE}$) and three dimensions (namely, $T_{CDE}$). The analysis of bilateral mutual information revealed decreasing trends that stabilize around small but nonzero values, indicating that the descriptors become increasingly and distinctly informative over time. Conversely, values of bilateral information above the 90\% confidence intervals are observed in those periods when a dominant branch used to classify publications in the field has not yet emerged. On the basis of the definition of the mutual information indication, a decrease in bilateral mutual information can be considered as a signal of differentiation among the $C$, $D$, and $E$ axes during the expansion and branching of a field due to new discoveries or technological advancements, and subsequent specialization in the relevant (sub-)fields. Thus, as an avenue of future research, this mutual information method provides a means to study the rate of specialization.

Despite the very different vocabulary spaces (that is, the three MeSH branches in use) and inception years, the analyses of trilateral mutual information revealed statistically significant (nonzero) negative mutual information values, $T_{CDE}(y)$, for each examined case-study, thereby rejecting the null hypothesis that $T_{CDE}=0$.\footnote{This is only the case when entropy is assessed on the basis of the full counts of the descriptors assigned to publications, i.e.\ when the whole publication-level information is preserved. When the other two threshold-based counting methods are considered (presence of a descriptor or the median value of the number of descriptors from a given branch within the list of descriptors assigned to publications by MEDLINE/PubMed), $T_{CDE}$ values did not significantly differ from zero, thus suggesting that the collection of MeSH descriptors is significantly more informative than just a count of descriptors on the basis of pre-selected thresholds. Binary counts cannot contain sufficient information to show the synergy effects because the values cannot be added, and no redundancy thus generated. (The Boolean ``1 OR 1'' remains ``1''.)} This indicates mutual redundancy shared among the three pairwise communication channels, corresponding to 'synergetic integration' or, in other words, emerging options, and therefore reduction of uncertainty. However, we found $T_{CDE}(y)$ contained within the 90\% confidence intervals for most of the observation period, but, when outside the confidence intervals, $T_{CDE}(y)$ often assumed more negative values indicating more 'synergetic integration' than expected from the background mixing levels of $C-D-E$.

The significant nonzero $T_{CDE}(y)$ further highlights that the governance of the (medical) innovation process (e.g.\ in translational medicine) should account for the key contributions from a variety of dimensions. In this paper, we conceptualized three dimensions in terms of \textit{supply}, \textit{demand}, and \textit{technological capabilities}, as well as the interactions among these dynamic communication channels. Left unaccounted, there is the possibility that policies can become 'locked-in' when the focus is only on two of the three dimensions, because the third may be spuriously structuring the configuration. For example, in the medical context, supporting the development of drugs for the treatment of a given disease and advance the understanding of the disease may not be a sufficient condition to generate innovation options. Indeed, along these lines, \cite{Yao2015} have used data-driven efforts to identify and quantify the institutional (funding) misalignment of biomedical supply and demand. Advances of \textit{technological capabilities} that enables new modalities of medical diagnosis and treatment along with the knowledge that accumulates with the use of these (through clinical practice) are also of critical importance \citep{Nelson2011}. In this regard, our analysis reveals that, in the \textit{supply-demand-technological capabilities} interplay, the \textit{technological capabilities} dimension was an important driving force for the three case-studies of medical areas we examined. This may suggest that this dimension can potentially function as control and support system for the governance of certain medical innovations.

The $T_{CED}$ indicator we developed, measuring the mutual information among \textit{supply}, \textit{demand}, and \textit{technological capabilities}, attempts to fill an important gap in science policy and innovation studies literature consisting of a missing link between the extensive conceptual efforts made on the role of uncertainty in technological change \citep[e.g.][]{Freeman1987} and the lack of indicators capable of providing a quantification of the uncertainty \citep{Rotolo2015}. From a governance perspective, such an indicator assumes even more importance considering its potential of informing the policy-making process about innovation uncertainty. Also, our study identifies the value in developing and maintaining classification systems that are capable of generating high-quality data, which are useful for increasing our quantitative understanding of the scientific knowledge order and, moreover, capable of keeping pace with the rapid change of the innovation process itself \citep{Griliches1994}. Indeed, the MeSH classification, for example, provides data that can function beyond their institutional context, i.e.\ the MEDLINE/PubMed article retrieval database. 

Our study presents some limitations that are worth discussing. First, we measured uncertainty in terms of mutual information between \textit{supply}, \textit{demand}, and \textit{technological capabilities}. Yet, different forms of uncertainty exist and these can be associated with different levels of 'ignorance' \citep{Stirling2007a} and hidden variables (i.e.\ additional dimensions). Future studies could further explore on the link between different forms of uncertainty and their operationalization. Second, we built knowledge representations of \textit{supply}, \textit{demand}, and \textit{technological capabilities} on the basis of publication data and the MeSH classification. Yet, these representations are focused on three branches of the classification, which provide a certain set of perspectives on the communication process in the medical context, as well biased toward medical research. Additional insights may come from the analysis of the mutual information among other branches of the classification as well as on the use of classification systems (e.g.\ patent technological classifications) that can provide a more comprehensive (proprietary in additional to academic) representation of medical applications. Finally, the presence of statistical regularities featuring relatively stable descriptive parameters, in addition to the common patterns of bilateral and trilateral mutual information we found across the examined case-studies, suggest that the growth of a research domain's communication system may be driven by fundamental underlying organizational processes that perhaps do not depend strongly on the specific details of a research area \citep[e.g.][]{Amaral1998}. This points to future research avenues, towards a systematic analysis of additional medical areas in order to generalize our findings. 

In summary, we decomposed the medical innovation process in terms of its functional components that were defined as \textit{supply}, \textit{demand}, and \textit{technological capabilities}. The resulting triple helix model of \textit{supply-demand-technological capabilities} enabled us to shed light on the bi- and tri-lateral interactions among these co-evolving selection environments and to specify an indicator of uncertainty that underlies the medical innovation process. The availability of knowledge representations of this process, in the form of high-quality (in breadth and depth) longitudinal data, enabled us to provide novel quantitative insights into three (Nobel Prize) research areas which capture the profound social and technological impact of biomedical innovation. Our main finding is the statistically significant nonzero values observed for $T_{CED}$, with negative values representing 'synergetic integration' across the three medical innovation channels we analyzed. As such, this important triple-interaction represents a key feature, neglected by linear and two-dimensional (e.g. \textit{supply-demand}) models, that should not be neglected by theories and policies developed in contexts where technological capabilities are essential. Moreover, within this three dimensional $C-D-E$ model summarized in \Figref{results}, we identified the demand-technology ($C-E$) pairwise channel to be the strongest relational link.

\section*{Acknowledgements}
We are grateful for constructive comments on a previous draft by two referees from the journal and two referees from the SPRU Working Paper Series. AMP acknowledges support from the Italian Ministry of Education for the National Research Project (PNR) ``CRISIS Lab''. AMP and LL acknowledge EU COST Action TD1210 ``KnowEscape''.  DR acknowledges the support of the People Programme (Marie Curie Actions) of the European Union's Seventh Framework Programme (FP7/2007-2013) (award PIOF-GA-2012-331107 - \href{http://www.danielerotolo.com/netgenesis}{``{\color{blue}NET-GENESIS: Network Micro-Dynamics in Emerging Technologies}}''). DR and LL acknowledge the support of the UK Economic and Social Research Council (award RES-360-25-0076, ``Mapping the Dynamics of Emergent Technologies'').

\newpage
\singlespace
\bibliographystyle{apalike}
\bibliography{/Users/danielerotolo/Dropbox/References/bibtex_references/library.bib}

\end{document}